\documentclass[onecolumn,11pt]{IEEEtran}
\usepackage{amsfonts,amsmath,amssymb,amsthm}
\usepackage[ruled,lined,linesnumbered]{algorithm2e}
\usepackage{amsbsy,caption}
\usepackage[para]{threeparttable}
\usepackage{graphicx,multirow,bm}
\usepackage{fancybox}
\usepackage{scalefnt}
\usepackage[noadjust]{cite}
\usepackage{tikz}
\usepackage{color}
\makeatletter
\newtheoremstyle{mythm}{3pt}{3pt}{}{16pt}{\bfseries}{:}{.5em}{}
\theoremstyle{mythm}
\newtheorem{theorem}{Theorem}
\newtheorem{example}{Example}

\newtheorem{remark}{Remark}

\newtheorem{corollary}{Corollary}
\newtheorem{lemma}{Lemma}

\newcommand{\cB}{\mathcal{B}}
\newcommand{\cC}{\mathcal{C}}
\newcommand{\cD}{\mathcal{D}}

\newcommand{\cP}{\mathcal{P}}

\newcommand{\cS}{\mathcal{S}}

\newcommand{\F}{\mathbb{F}}
\newcommand{\Tr}{\mathrm{Tr}}
\newcommand{\Norm}{\mathrm{Norm}}

\renewcommand{\leq}{\leqslant}

\renewcommand{\geq}{\geqslant}

\begin{document}

\title{Subfield Codes of Several Few-Weight Linear Codes  Parametrized by  Functions and Their Consequences}
\author{Li Xu, Cuiling Fan, Sihem Mesnager, Rong Luo, Haode Yan
\thanks{Li Xu, Cuiling Fan, Rong Luo and Haode Yan are with the School of Mathematics, Southwest Jiaotong University, Chengdu, 610031, China (e-mail: xuli1451@163.com; zzc@swjtu.edu.cn; luorong@swjtu.edu.cn; hdyan@swjtu.edu.cn).}
\thanks{Sihem Mesnager is with the Department of Mathematics, University of Paris VIII, 93526 Saint-Denis, France, University of Paris XIII, CNRS, UMR 7539 LAGA, Sorbonne Paris Cit$\acute{\mathrm{e}}$, 93430 Villetaneuse, France, and Telecom Paris, Polytechnic Institute of Paris, 91120 Palaiseau, France. (e-mail: smesnager@univ-paris8.fr).}
}

\maketitle

\begin{abstract}
Subfield codes of linear codes over finite fields have recently received much attention. Some of these codes are optimal and have applications in secret sharing, authentication codes and association schemes.
In this paper, the $q$-ary subfield codes $\cC_{f,g}^{(q)}$ of six different families of a linear code $\cC_{f,g}$  parametrized by two functions $f, g$ over a finite field $\mathbb{F}_{q^m}$ are considered and studied, respectively.
The parameters and  (Hamming) weight distribution of   $\cC_{f,g}^{(q)}$ and their punctured codes $\bar{\cC}_{f,g}^{(q)}$ are explicitly determined. The parameters of the duals of these codes are also analyzed.
Some of the resultant $q$-ary codes $\cC_{f,g}^{(q)}$, $\bar{\cC}_{f,g}^{(q)}$ and their dual codes are optimal, and some have the best-known parameters.
The parameters and weight enumerators of the first two families of linear codes $\cC_{f,g}$ are also settled, among which the first family is an optimal two-weight linear code meeting the Griesmer bound, and the dual codes of these two families are almost maximum distance separable codes (MDS codes).
As a byproduct of this paper, a family of $[2^{4m-2},2m+1,2^{4m-3}]$ quaternary Hermitian self-dual codes are obtained with $m \geq 2$.
As an application, we show that three families of the derived linear codes give rise to several infinite families of $t$-designs  ($t\in\{2,3\}$).


\end{abstract}


\section{Introduction}
\IEEEPARstart{L}{et} $p$ be a prime, $q=p^l$ for a positive integer $l$. Let $\F_q$ be the finite field with $q$ elements.
A $q$-ary $[n, k, d]$ linear code is a $k$-dimensional subspace of $\F^n_q$ with minimum Hamming distance $d$. 
An $[n, k, d]$ linear code ${\cal C}$ is said to be optimal if no $[n, k, d']$ code with $d' > d$ exists.
The generator matrix of ${\cal C}$ is a $k \times n$ matrix $G$ whose rows form a basis of ${\cal C}$ as an $\F_q$-vector space, since ${\cal C}=\{{\bf x} G:{\bf x}\in \F_q^k\}$.
The dual code of $\mathcal C$ is defined as $\mathcal C^\perp=\{{\bm x}\in \mathbb{F}_q^n:  {\bm x}\cdot{\bm c}=0~{\rm for~all}~{\bm c}\in \mathcal C\},$ where $ {\bm x}\cdot {\bm c}=\sum_{i=1}^nx_ic_i.$
For some $i \in \{1,2, \cdots, n\}$, puncture $\cC$ by deleting the same coordinate $i$ in each codeword, and denote the punctured code by $\bar{\cC}$. If $G$ is a generator matrix for $\cC$, then a generator matrix for $\bar{\cC}$ is obtained from $G$ by deleting column $i$ (and omitting a zero or duplicate rows that may occur).

Let $A_i$ denote the number of codewords with Hamming weight $i$ in a code $\cC$ of length $n$.
The weight enumerator of $\cC$ is defined by $1 + A_1z + A_2z^2 + \cdots + A_nz^n$.
The sequence $(A_0=1, A_1, A_2,\cdots, A_n)$ is called the weight distribution of the code.
The study of the weight distribution of a linear code is important both in theory and applications due to the following (\cite{Klove2007}):
\begin{itemize}
 \item The weight distribution gives the minimum distance of the code and, thus, the error-correcting capability.
 \item The weight distribution of a code allows the computation of the error probability of error detection and correction with respect to some error detection and error correction algorithms.
\end{itemize}
A code $\cC$ is said to be a $t$-weight code if the number of nonzero $A_i$ in the sequence $(A_1, A_2,\cdots, A_n)$ is equal to $t$. 
Recall that MDS  $[n,k,d]$ linear codes are those whose well-known singleton bound  $k + d \leq n + 1$ on their parameters is achieved with equality (that is, $k + d=n + 1$).  In addition to their theoretical importance in coding theory, they have gained great public interest thanks to their important applications in distributed storage systems and error-correcting communication, particularly channels susceptible to burst errors.  If $d = n-k$, then the code is called an almost MDS code.

The construction of linear codes with few weights is also a meaningful research
because they have important applications in secret sharing \cite{Anderson1998,Ding2005-1}, association schemes \cite{Calderbank1984}, strongly regular graphs \cite{Calderbank1986} and authentication codes \cite{Ding2005}.
In finite geometry, hyperovals in PG$(2, 2^m)$ and conics in PG$(2, q)$ are the same as those MDS codes with two or three weights \cite{Ding2018}, maximal $(n,h)$-arcs in PG$(2, 2^m)$ and ovoids in PG$(3, q)$ are also the same as a particular type of two-weight codes \cite{Ding2018}.

For more information on few-weight linear codes, the reader is referred to \cite{Ding2015,Ding2016-1,Ding2007,Ding2014,Heng2015,Li2017,Luo2018,Mesnager2020,Mesnager2021,Tan2018,Tang2016,Tang2016-2,Zhou2013,Zhou2013-2,Zhou2016} and the references therein.
One of the highlights of the present paper is to determine the weight distributions of some linear codes and design new ones with few weights.

Subfield codes were first studied in \cite{Canteaut2000,Carlet1998} to construct the linear codes with good parameters over $\F_q$ from linear codes over $\F_{q^m}$.
Recently, some primary results about subfield codes were derived, and the subfield codes of ovoid codes were studied in \cite{Ding2019}. It was demonstrated that the subfield codes of ovoid codes are desirable (\cite{Ding2019}).
The subfield codes of some hyperoval codes and conic codes were also studied in \cite{Heng2019}, and these results were later extended in \cite{Wang2020,Wang2021}.
In \cite{Heng2020}, optimal binary linear codes were constructed from maximal arcs.
Meanwhile, in \cite{Heng2022}, they presented the subfield codes of $[q+1,2,q]$ MDS codes with different forms of generator matrix and some optimal linear codes are obtained.
The subfield codes of some cyclic codes were also studied in \cite{Heng2020-2,Tang2022}. More subsequent papers on subfield codes include \cite{Zheng2020,Wang2021,Xiang2021,Wu2020,Wu2022}.
As shown in these mentioned papers, subfield codes of linear codes are a significant study object in coding theory because some codes are optimal (or almost optimal) and always have few weights.

Let $f(x)$ and $g(x)$ be two different functions from $\F_{q^m}$ to itself, where $m$ is a positive integer. 
Denote
$$\cS_{f,g}=\{(1,x,y) : (x, y) \in  \F_{q^m}^2,~f(x) + g(y)=0\}$$
and
$$G_{\cS_{f,g}}=\left(\begin{array}{l}
 1\\
 x\\
 y
\end{array}\right)_{(x,y) \in  \F_{q^m}^2,~f(x) + g(y)=0,}$$
which is a $3 \times (\#\cD)$ matrix over $\F_{q^m}$, where $\#\cD$ is the cardinality of set $\cD=\{(x,y) \in  \F_{q^m}^2 : f(x) + g(y)=0\}.$
We construct a linear code $\cC_{f,g}$ over $\F_{q^m}$ with generator matrix
\begin{equation}\label{G}
G_{f,g}=\left(\begin{array}{ll}
0 & \\
1 & G_{\cS_{f,g}}\\
0 &
\end{array}\right).
\end{equation}

Some known linear codes can be obtained by selecting appropriate functions $f(x)$ and $g(y)$.
It turns out that
\begin{itemize}
\item ${\cal C}_{f,g}$ is a punctured hyperoval code when $q$ is even, $f(x)=x$ and $g(y)$ is an oval polynomial (\cite{Hirschfeld1998});
 \item ${\cal C}_{f,g}$ is a conic code when $q$ is odd, $f(x)=x$ and $g(y)=y^2$ (\cite{Hirschfeld1998}).
\end{itemize}

Throughout this paper, we use $\Tr_{q^m/q}$ and $\Norm_{q^m/q}$ to denote the trace and norm functions from $\F_{q^m}$ onto $\F_{q}$, respectively, which are defined by
$$\Tr_{q^m/q}(x)= x + x^q + \cdots + x^{q^{m-1}},$$
$$\Norm_{q^m/q}(x)= x^{(q^m-1)/(q-1)},$$
where $x \in \F_{q^m}$. Let $A(x)$ be an almost bent function and $B(x)$ be a Boolean bent function.
In this paper, we focus on six families of linear codes ${\cal C}_{f,g}$, where $f$ and $g$ are distinct and selected from the set of functions $\{\Tr_{q^m/q}(x), \Tr_{q^m/q}(x^2), \Norm_{q^m/q}(x), \Tr_{2^m/2}\left(A(x)\right), B(x)\}$.

When $m=2$, $f(x)=\Tr_{q^2/q}(x)$ and $g(x)=\Norm_{q^2/q}(x)$, we show that $\cC_{f,g}$ is an optimal two-weight code achieving the Griesmer bound, and its dual is an almost MDS code. For general integer $m$, when $f(x)=\Tr_{q^m/q}(x)$ and $g(x)=\Tr_{q^m/q}(x^2)$, $\cC_{f,g}$ is shown to have four or five weights, its dual is also an almost MDS code.
The subfield codes and the punctured code of the subfield codes of these two families of few-weight codes are also studied. These resultant codes are $q$-ary few-weight codes, some of them and their dual codes are optimal, and some have the best-known parameters.
Especially when $q=2$, the binary subfield codes and the punctured codes of the subfield codes of four different families of $\cC_{f,g}$ are studied with $\left(f(x), g(x)\right) \in \left\{\left(\Tr_{2^m/2}(x),\Tr_{2^m/2}(A(x))\right), \left(\Tr_{2^m/2}(A_1(x)),\Tr_{2^m/2}(A_2(x))\right), \left(\Tr_{2^m/2}(x), B(x)\right),\left(B_1(x), B_2(x)\right) \right\}$.
Seven families of few-weight binary linear codes are presented, and most of their dual codes are optimal concerning the sphere packing bound.

The rest of this paper is structured as follows.
Section \ref{Pre} recalls some notations and basics of characters, linear codes and some special functions, which will be used in subsequent sections.
Section \ref{f,g} studies the $q$-ary subfield codes $\cC_{f,g}^{(q)}$ and the punctured code $\bar{\cC}_{f,g}^{(q)}$ of the subfield codes of six different families of $\cC_{f,g}$ respectively.
The parameters and weight enumerators of the first two families of linear codes $\cC_{f,g}$ are also determined.
Some of the resultant $q$-ary few-weight codes and their dual codes are optimal, and some have the best-known parameters. As a byproduct, a family of $[2^{4m-2},2m+1,2^{4m-3}]$ quaternary Hermitian self-dual
code are obtained, where $m \geq 2$.
Section \ref{design} presents several infinite families of $2$-designs or $3$-designs with some of the codes presented in this paper.
Section \ref{conclusion} summarizes this paper.

\section{Preliminaries}\label{Pre}

This section presents some basic notations, definitions, and necessary auxiliary results for the subsequent sections. We fix the following notations unless otherwise stated. It specifically briefly introduces some known results about characters and linear codes over finite fields and special functions on $\F_{2^m}$, which will be used later in this paper.

\subsection{Characters Over Finite Fields}

Let $\F_q$ be a finite field with $q$ elements, where $q$ is a power of a prime $p$. 
Denote by $\zeta_p$ the primitive $p$-th root of complex unity. The additive character of $\F_q$ is defined as a homomorphism $\chi$ from $\F_q$ into the complex unit group such that $\chi (x + y) = \chi (x)\chi (y)$ for $x, y \in \F_q$. For any $a \in \F_q$, the function defined by
$$\chi_a(x) = \zeta_p^{\Tr_{q/p}(ax)},~ x \in \F_q$$
is an additive character of $\F_q$. In addition, $\{\chi_a : a \in \F_q\}$ is a group containing all the additive character of $\F_q$. When $a=0$, we obtain the trivial additive character $\chi_0$, for which $\chi_0(x)=1$ for all $x \in \F_q$. When $a=1$, $\chi_1$ is called the canonical additive character of $\F_q$. It's obvious that $\chi_a(x) = \chi_1(ax)$.
A crucial property of the additive characters, called the orthogonality (\cite{Lidl1997}), is given as follows:
$$
\sum_{x \in \F_q} \chi_1(a x)= \begin{cases}q & \text { for } a=0, \\ 0 & \text { for } a \in \F_q^*,\end{cases}
$$
where $\F_q^*=\F_q \backslash \{0\}.$

A character of the multiplicative group $\F_q^*$ is defined as a homomorphism $\psi$ from $\F_q^*$ into the complex unit group such that $\psi(xy) = \psi(x) \psi(y)$ for all $x, y \in \F_q^*$. $\psi$ is also called multiplicative character of $\F_q$.
Let $g$ be a fixed primitive element of $\F_q$. For each $j=0,1,\cdots,q-2$, the function
$$
\psi_j\left(g^k\right)=\zeta_{q-1}^{j k} \text { for } k=0,1, \cdots, q-2
$$
defines a multiplicative character of $\F_q$, and every multiplicative character of $\F_q$ is obtained in this way.
No matter what $g$ is, the character $\psi_0$ will always represent the trivial multiplicative character, which satisfies $\psi_0(x)=1$ for all $x \in \F_q^*$. $\eta:=\psi_{(q-1) / 2}$ is called the quadratic character of $\F_q$. The orthogonality relation of multiplicative characters is given by
$$
\sum_{x \in \F_q^{*}} \psi_j(x)=\begin{cases}
q-1 & \text { for } j=0, \\
0 & \text { for } j \neq 0 .
\end{cases}
$$

For an additive character $\chi$ and a multiplicative character $\psi$ of $\F_q$, the Gauss sum $G(\psi, \chi)$ over $\F_q$ is defined by
$$
G(\psi, \chi)=\sum_{x \in \F_q^{*}} \psi(x) \chi(x) .
$$
We call $G(\eta, \chi)$ the quadratic Gauss sum over $\F_q$ for nontrivial $\chi$. The value of the quadratic Gauss sum is known and documented below.

\begin{lemma}[\cite{Lidl1997}]\label{Gauss}
 Let $q=p^l$ with an odd prime $p$ and a positive integer $l$. Let $\chi$ be the canonical additive character of $\F_q$. Then
$$
\begin{aligned}
G(\eta, \chi) &=\begin{cases}
(-1)^{l-1} \sqrt{q} & \text { if } p \equiv 1 ~ \bmod ~4, \\
(-1)^{l-1}(\sqrt{-1})^l \sqrt{q} & \text { if } p \equiv 3 ~\bmod ~4,
\end{cases}\\
&=(-1)^{l-1}(\sqrt{-1})^{\left(\frac{p-1}{2}\right)^2 l} \sqrt{q}.
\end{aligned}
$$
\end{lemma}

Let $\chi$ be a nontrivial additive character of $\F_q$ and let the polynomial $f \in \F_q[x]$ be
of positive degree. The character sums of the form
$$\sum\limits_{c \in \F_q} \chi\left(f(c)\right)$$
are sometimes referred to as Weil sums. The problem of evaluating such character sums explicitly
is difficult.
One usually has to be satisfied with estimates for the absolute value of the sum.
These character sums can be treated (see, e.g.  \cite{Lidl1997}).

When $f$ is a quadratic polynomial, and $q$ is odd, the Weil sums have an interesting relationship with quadratic Gauss sums, described in the following lemma.

\begin{lemma}[\cite{Lidl1997}]\label{character_odd}
 Let $\chi$ be a nontrivial additive character of $\F_q$ with $q$ odd, and let $f(x)=a_2 x^2+a_1 x+a_0 \in \F_q[x]$ with $a_2 \neq 0$. Then
$$
\sum_{c \in \F_q} \chi(f(c))=\chi\left(a_0-a_1^2\left(4 a_2\right)^{-1}\right) \eta\left(a_2\right) G(\eta, \chi).
$$
\end{lemma}

The Weil sums can also be evaluated explicitly if $f$ is a quadratic polynomial and $q$ is even.

 \begin{lemma}[\cite{Lidl1997}]\label{character_even}
Let $\chi_b$, $b \in$ $\F_q^*$, be a nontrivial additive character of $\F_q$ with $q$ even, and let $f(x)=a_2 x^2+a_1 x+a_0 \in \F_q[x]$. Then
$$
\sum_{c \in \F_q} \chi_b(f(c))= \begin{cases}\chi_b\left(a_0\right) q & \text { if } a_2=b a_1^2, \\ 0 & \text { otherwise. }\end{cases}
$$
\end{lemma}




We consider now character sums involving only the quadratic character $\eta$ of $\F_q$, $q$ odd, and having quadratic polynomial arguments, that is, sums of the form
$$\sum\limits_{c \in \F_q} \eta\left(f(c)\right)$$
with $f(x)=a_2 x^2+a_1 x+a_0 \in \F_q[x]$.
The explicit formula is as follows.

\begin{lemma}[\cite{Lidl1997}]\label{eta}
Let $f(x)=a_2 x^2+a_1 x+a_0 \in \F_q[x]$ with $q$ odd and
$a_2 \neq 0$. Put $d = a_1^2-4a_0a_2$. Then
$$
\sum_{c \in \F_q} \eta(f(c))= \begin{cases}
-\eta(a_2) & \text { if } d\neq 0, \\
(q-1)\eta(a_2) & \text { if } d=0.
 \end{cases}
$$

\end{lemma}

Let $f \in \F_q[x]$ be a monic quadratic polynomial, and then the number of solutions of $f(x)=0$ in $\F_q$ is given by the following lemma.

\begin{lemma}[\cite{Lidl1997}]\label{tr}
 Let $a_0, a_1 \in\F_{q}$, where $q=p^l$. Define $N_{a_0,a_1}=\#\{x\in\F_{q}:x^2+a_1x+a_0=0\}$. Then $N_{a_0,a_1} \in \{0,1,2\}$. More precisely, for $p=2$,
$$ N_{a_0,a_1}=\begin{cases}
0  & \text { if } ~\Tr_{2^l/2}(\frac{a_0}{a_1^2})=1, \\
1  & \text { if } ~a_1=0, \\
2 & \text { if } ~\Tr_{2^l/2}(\frac{a_0}{a_1^2})=0;
\end{cases} $$
and for $p>2$,
$$ N_{a_0,a_1}=\begin{cases}
0  & \text { if } ~\eta(a_1^2-4a_0)=-1, \\
1  & \text { if } ~a_1^2-4a_0=0, \\
2 & \text { if } ~\eta(a_1^2-4a_0)=1.
\end{cases} $$
\end{lemma}

\subsection{Finite Projective Geometry}

The projective space of dimensional $r$ obtained from $\F_q$ will be denoted by PG$(r,q)$, where its points are the one-dimensional subspaces of $\F_q^{r+1}$ and its hyperplanes are the $r$-dimensional subspaces of $\F_q^{r+1}$.
Obviously any hyperplane can be defined to be ${\cal H}_{\bf u}=\{{\bf x} \in \F_q^{r+1}: {\bf x}\cdot{\bf u}=0\}$ for some nonzero vector ${\bf u}\in \F_q^{r+1}$. 

For a $q$-ary $[n,k,d]$ linear code ${\cal C}$ with a generator matix $G=\left({\bf g}_1\ldots{\bf g}_n\right)$, if we view the columns of $G$ to be projective points (maybe repeated) in PG$(k-1,q)$, and define $S=\{{\bf g}_1,\ldots,{\bf g}_n\}$ to be an multiset.
 The following lemma states a relationship between weights of codewords in ${\cal C}$ and hyperplanes in PG$(k-1,q)$.

\begin{lemma}[\cite{Ball2015}]\label{Weight-lem}
Let ${\bf u}$ be a nonzero vector of $\F_q^k$. The codeword ${\bf u}G$ has weight $w$ if and only if the hyperplane ${\cal H}_{\bf u}$ in PG$(k-1,q)$ contains $n-w$ points of $S$, i.e., $wt({\bf u}G)=n-\#({\cal H}_{\bf u}\cap S)$.
\end{lemma}

From this lemma, for any $0< i \leq n$, the number of codewords with weight $i$ in code ${\cal C}$ is
$$A_i=\#\{{\bf u}\in{\mathbb F}_q^k\setminus\{{\bf 0}\}:\#({\cal H}_{\bf u}\cap S)=n-i\}.$$
This property plays an important role in determining the weight distribution of linear code $\cC_{f,g}$ generated by (\ref{G}).

\subsection{The Subfield Codes of Linear Codes}

Given an $[n, k]$ linear code over $\F_{q^m}$. We construct a new $[n, k']$ code $\cC^{(q)}$
over $\F_q$ as follows.
Let $G$ be a generator matrix of $\cC$. Take a basis of $\F_{q^m}$ over $\F_q$. Represent each entry of $G$ as an $m \times 1$ column vector of $\F^m_q$ with respect to this basis,
and replace each entry of $G$ with the corresponding $m \times 1$ column vector of $\F^m_q$.
In this way, $G$ is modified into a $km \times n$ matrix over $\F_q$, which generates  the new subfield code $\cC^{(q)}$ over $\F_q$ with length $n$.
It is known that the subfield code $\cC^{(q)}$ is independent of both the choice of the basis of $\F_{q^m}$ over $\F_q$ and the choice of the generator matrix $G$ (\cite{Ding2019}).

By definition, the dimension $k'$ of $\cC^{(q)}$ satisfies $k' \leq mk$.
The relationship between the minimal distance of $\cC^{\perp}$ and that of $\cC^{(q)\perp}$ is given as follows.

\begin{lemma}[\cite{Ding2019}]\label{distance}
 The minimal distance $d^{\perp}$ of $C^{\perp}$ and the minimal distance $d^{(q)\perp}$ of $C^{(q)\perp}$ satisfy
$$d^{(q)\perp} \geq d^{\perp}.$$
\end{lemma}

The trace representation of the $q$-ary subfield code $\mathcal{C}^{(q)}$ of $\cC$ is given in the next lemma.

\begin{lemma}[\cite{Ding2019}]\label{trace-repre}
 Let $\mathcal{C}$ be an $[n, k]$ linear code over $\mathbb{F}_{q^m}$. Let $G=\left[g_{i j}\right]_{1 \leq i \leq k, 1 \leq j \leq n}$ be a generator matrix of $\mathcal{C}$. Then the trace representation of the subfield code $\mathcal{C}^{(q)}$ is given by
$$
\mathcal{C}^{(q)}=
\quad\left\{\left(\operatorname{Tr}_{q^m/q}\left(\sum_{i=1}^k a_i g_{i 1}\right), \cdots, \operatorname{Tr}_{q^m/q}\left(\sum_{i=1}^k a_i g_{i n}\right)\right): a_1, \ldots, a_k \in \mathbb{F}_{q^m}\right\}.
$$
\end{lemma}

\begin{remark}
It follows from this lemma that the $q$-ary subfield code $\mathcal{C}^{(q)}$ of a linear code $\cC$ over $\mathbb{F}_{q^m}$ is, in fact, the trace code $\Tr_{q^m/q}(\cC)$ of $\cC$, which is different from the subfield subcode well studied in the literature.
\end{remark}


\subsection{Pless Power Moments and Two Bounds for Linear Codes}
We shall need the Pless power moments for linear codes to study the minimal distances of the dual codes of some codes. Let $\cC$ be a $q$-ary $[n, k]$ code with weight distribution $(1, A_1,\cdots, A_n)$, we denote by $(1, A_1^{\perp} ,\cdots, A_n^{\perp} )$ the weight distribution of its dual code.  The first four Pless
power moments on these two weight distributions are given as follows \cite{Huffman}:

\begin{align*}
&\sum_{j=0}^n A_j=q^k,\\
&\sum_{j=0}^n j A_j=q^{k-1}\left(q n-n-A_1^{\perp}\right),\\
&\sum_{j=0}^n j^2 A_j=q^{k-2}(q-1) n(q n-n+1)-q^{k-2}\left((2 q n-q-2 n+2) A_1^{\perp}+2 A_{2}^{\perp}\right),
\end{align*}
\begin{align*}
\sum_{j=0}^n j^3 A_j= & ~q^{k-3}\left[(q-1) n\left(q^2 n^2-2 q n^2+3 q n-q+n^2-3 n+2\right)\right.\\
                     &\left.- \left(3 q^2 n^2-3 q^2 n-6 q n^2+12 q n+q^2-6 q+3 n^2-9 n+6\right) A_1^{\perp}
+6(qn-q-n+2)A_2^{\perp}-6A_3^{\perp}\right].
\end{align*}

We will also need the following two classical bounds for linear codes (\cite{Huffman}).

\begin{lemma}[Griesmer bound] Let $\cC$ be an $[n,k,d]$ linear code over $\F_q$ with $k\geq 1$. Then
$$
n \geq \sum_{i=0}^{k-1}\left\lceil\frac{d}{q^i}\right\rceil,
$$
where $\lceil \cdot \rceil$ denotes the ceiling function.
\end{lemma}

\begin{lemma}[Sphere packing bound] Let $\cC$ be an $[n,k,d]$ linear code over $\F_q$. Then
$$
q^n  \geq q^k \sum_{i=0}^{t} \binom{n}{i}(q-1)^i,
$$
where $t=\left\lfloor\frac{d-1}{2}\right\rfloor$ and $\lfloor \cdot \rfloor$ denotes the floor function.
\end{lemma}

\subsection{(Almost) Bent Functions}

Let $f(x)$ be a function from $\F_{2^m}$ to $\F_{2^t}$.
The Walsh transform of $f(x)$ at $(a, b)\in \F_{2^t}^* \times \F_{2^m}$ is defined as
$$W_f(a, b)= \sum\limits_{x \in \F_{2^m}} (-1)^{\Tr_{2^t/2}(af (x))+\Tr_{2^m/2}(bx)}. $$
Specially, when $t=m$, the Walsh transform of $f$ is then of the form
$$W_f(a, b)= \sum\limits_{x \in \F_{2^m}} (-1)^{\Tr_{2^m/2}(af (x)+bx)}. $$
A function $f : \F_{2^m}\rightarrow \F_{2^m}$ is said to be an almost bent function if $W_f(a, b) = 0$ or $\pm 2^{\frac{m+1}{2}}$ for any pair $(a, b) \in \F_{2^m}^* \times \F_{2^m}$. Almost bent functions exist only when $m$ is odd (\cite{Carlet1998}).

When $t=1$, the Walsh transform is then of the form
$$W_f(1, b)=W_f(b)= \sum\limits_{x \in \F_{2^m}} (-1)^{f (x)+\Tr_{2^m/2}(bx)}. $$
A function $f : \F_{2^m}\rightarrow \F_{2}$ is said to be a Boolean bent function if $W_f(b) = \pm 2^{\frac{m}{2}}$ for any $b \in \F_{2^m}$. Boolean bent functions exist only for even $m$ (\cite{Rothaus1976}).


Almost bent and Boolean bent functions play important roles in coding theory, cryptography, sequences, and combinatorics (see, \cite{Carlet2021,CarletMesnagerDCC2016, Mesnager-Book}). Many good linear codes over finite fields have been constructed with almost bent and Boolean bent functions \cite{Carlet1998,Ding2015, Ding2016-1, Li2014-3, Mesnager2017, Mesnager2021, Tang2016}. In the present paper, they will be used to construct optimal binary codes with respect to the sphere packing bound.

\section{The subfield codes of $\cC_{f,g}$}\label{f,g}
Let $\cC_{f,g}$ be the linear code over $\F_{q^m}$ generated by (\ref{G}),
where $f$ and $g$ are two different functions from $\F_{q^m}$ to $\F_q$.
Due to Lemma \ref{trace-repre},
the trace representation of the subfield code $\mathcal{C}_{f,g}^{(q)}$ is given by
$$\cC_{f,g}^{(q)} =\left\{ \textbf{c}_{a,b,c}: a \in \F_{q}, b ,c \in \F_{q^m}\right\},$$
where
$$\textbf{c}_{a,b,c}=\left( \Tr_{q^m/q}(b),~ (a+\Tr_{q^m/q}(bx+cy))_{(x,y) \in \cD }\right),~\cD=\{(x,y) \in \F_{q^m}^2: f(x) + g(y)=0\}.$$

Let $\chi$ and $\chi'$ be the canonical additive characters of $\F_q$ and $\F_{q^m}$, respectively, for the rest of the paper.

\begin{lemma}\label{length}
The code length of the code $\mathcal{C}_{f,g}^{(q)}$ is
$$n=1+q^{2m-1}+\frac{1}{q}\sum\limits_{z\in \F_q^*} \sum\limits_{x \in \F_{q^m}}\chi(zf(x)) \sum\limits_{y \in \F_{q^m}}\chi(zg(x)).$$
\end{lemma}
\begin{IEEEproof}
By the orthogonality of group characters, we have
 \[\begin{aligned}
\#\cD &= \frac{1}{q}\sum\limits_{(x,y) \in \F_{q^m}^2} \sum\limits_{z\in \F_q} \chi\left(z(f(x)+g(y))\right)\\
 &=q^{2m-1} + \frac{1}{q}\sum\limits_{z\in \F_q^*} \sum\limits_{(x,y) \in \F_{q^m}^2} \chi\left(zf(x)+ zg(y)\right).
  \end{aligned}\]
Then $n=1+ \#\cD = 1+q^{2m-1}+\frac{1}{q}\sum\limits_{z\in \F_q^*} \sum\limits_{x \in \F_{q^m}}\chi(zf(x)) \sum\limits_{y \in \F_{q^m}}\chi(zg(x)).$
\end{IEEEproof}

We define a function $\delta(x)$ from $\F_{q^m}$ to $\{0,1\}$ as
\begin{equation}\label{delta}
 \delta(x)=
   \begin{cases}
  0  & \text { if } \Tr_{q^m/q}(x)=0,\\
  1  & \text { if } \Tr_{q^m/q}(x)\neq 0. \\
    \end{cases}
\end{equation}


\begin{lemma}\label{weight}

For any $a \in \F_q$, $b,c \in \F_{q^m}$, the weight of a codeword $\textbf{c}_{a,b,c}$ in $\cC_{f,g}^{(q)}$ is given by
$$
  wt(\textbf{c}_{a,b,c})
  =\begin{cases}
  0 & \text { if } a=b=c=0,\\
  \#\cD  & \text { if } a\neq 0, b=c=0, \\
  \delta(b) +  \frac{q-1}{q}(\#\cD) - \frac{1}{q^2}\Upsilon_{a,b,c}  & \text { if $b$ and $c$ are not all 0},\\
    \end{cases}
$$
where $\Upsilon_{a,b,c}=\sum\limits_{z\in \F_q^{*}} \chi(za) \sum\limits_{w \in \F_q^*}\sum\limits_{(x,y) \in \F_{q^m}^2}\chi(w f(x) + wg(y))\chi'(zbx+zcy)$.
\end{lemma}
\begin{IEEEproof}
Denote
$$N_{a,b,c}= \# \{ (x,y) \in \cD : a+\Tr_{q^m /q}(bx+cy)=0\}.$$
By the orthogonality relation of additive characters and the transitivity of trace functions, we have

 \[\begin{aligned}
 qN_{a,b,c}&= \sum\limits_{(x,y) \in \cD} \sum\limits_{z\in \F_q} \chi\left(z\left(a+\Tr_{q^m /q}(bx+cy)
\right)\right)\\
 &=\#\cD + \sum\limits_{(x,y) \in \cD} \sum\limits_{z\in \F_q^{*}} \chi\left(za+\Tr_{q^m /q}(zbx+zcy)\right)\\
 &=\#\cD + \sum\limits_{z\in \F_q^{*}} \chi(za) \sum\limits_{(x,y) \in \cD} \chi'(zbx+zcy)\\
 &=\#\cD + \sum\limits_{z\in \F_q^{*}} \chi(za) \sum\limits_{(x,y) \in \F_{q^m}^2} \left( \frac{1}{q}\sum\limits_{w \in \F_q}\chi\left(w\left(f(x) + g(y)\right)\right) \right) \chi'(zbx+zcy)\\
 &=\#\cD + \frac{1}{q}\sum\limits_{z\in \F_q^{*}} \chi(za) \sum\limits_{(x,y) \in \F_{q^m}^2} \chi'(zbx+zcy)
  + \frac{1}{q}\sum\limits_{z\in \F_q^{*}} \chi(za) \sum\limits_{w \in \F_q^*}\sum\limits_{(x,y) \in \F_{q^m}^2}\chi(w f(x) + wg(y))\chi'(zbx+zcy),\\
 \end{aligned}\]
 where $\#\cD$ is shown in the proof of Lemma \ref{length}. We deduce that
  $$
 N_{a,b,c}=\begin{cases}
   \#\cD & \text { if } a=b=c=0,\\
  0  & \text { if } a\neq 0, b=c=0, \\
  \frac{1}{q}(\#\cD) + \frac{1}{q^2}\sum\limits_{z\in \F_q^{*}} \chi(za) \sum\limits_{w \in \F_q^*}\sum\limits_{(x,y) \in \F_{q^m}^2}\chi\left(w f(x) + wg(y)\right)\chi'(zbx+zcy)  & \text { if $b$ and $c$ are not all 0}. \\
    \end{cases}
 $$
Hence,
  \[\begin{aligned}
  wt(\textbf{c}_{a,b,c}) &=\delta(b)+ \#\cD - N_{a,b,c}\\
  &=\begin{cases}
  0 & \text { if } a=b=c=0,\\
  \#\cD  & \text { if } a\neq 0, b=c=0, \\
  \delta(b) +  \frac{q-1}{q}(\#\cD) - \frac{1}{q^2}\Upsilon_{a,b,c}  & \text { if $b$ and $c$ are not all 0}, \\
    \end{cases}
   \end{aligned}\]
  where $\Upsilon_{a,b,c}=\sum\limits_{z\in \F_q^{*}} \chi(za) \sum\limits_{w \in \F_q^*}\sum\limits_{(x,y) \in \F_{q^m}^2}\chi\left(w f(x) + wg(y)\right)\chi'(zbx+zcy)$.
\end{IEEEproof}

Let $\bar{\cC}_{f,g}^{(q)}$ be the code $\cC_{f,g}^{(q)}$ punctured on the first coordinate. That is,
\begin{equation}\label{bar}
\bar{\cC}_{f,g}^{(q)} =\left\{ \bar{\textbf{c}}_{a,b,c}: a \in \F_{q}, b ,c \in \F_{q^m}\right\},
\end{equation}
where
$$\bar{\textbf{c}}_{a,b,c}=\left(a+\Tr_{q^m /q}(bx+cy)\right)_{(x,y) \in \cD },~\cD=\{(x,y) \in \F_{q^m}^2: f(x) + g(y)=0\}.$$

From the above two lemmas, the following conclusions can be easily drawn.

\begin{lemma}\label{weight-2}
The code length of $\bar{\mathcal{C}}_{f,g}^{(q)}$ is
$$n= \#D= q^{2m-1}+\frac{1}{q}\sum\limits_{z\in \F_q^*} \sum\limits_{x \in \F_{q^m}}\chi(zf(x)) \sum\limits_{y \in \F_{q^m}}\chi(zg(x)).$$
For any $a \in \F_q$, $b,c \in \F_{q^m}$, the weight of a codeword $\bar{\textbf{c}}_{a,b,c}$ in $\bar{\cC}_{f,g}^{(q)}$ is given by
$$
  wt(\textbf{c}_{a,b,c})
  =\begin{cases}
  0 & \text { if } a=b=c=0,\\
  \#\cD  & \text { if } a\neq 0, b=c=0, \\
  \frac{q-1}{q}(\#\cD) - \frac{1}{q^2}\Upsilon_{a,b,c}  & \text { if $b$ and $c$ are not all 0},\\
    \end{cases}
$$
where $\Upsilon_{a,b,c}$ is shown in Lemma \ref{weight}.
\end{lemma}

Now we consider six different pairs of functions $f(x)$ and $g(x)$ from the set $\{\Tr_{q^m/q}(x), \Tr_{q^m/q}(x^2), \Norm_{q^m/q}(x),$ $ \Tr_{2^m/2}\left(A(x)\right), B(x)\}$, where $A(x)$ and $B(x)$ are almost bent and Boolean bent functions, respectively.

\subsection{$f_1(x)=\Tr_{q^2/q}(x)$ and $g_1(y)=\Norm_{q^2/q}(y)$}\label{norm}

In this subsection, let $m=2$ and $f_1(x)=\Tr_{q^2/q}(x)$, $g_1(y)=\Norm_{q^2/q}(y)$. For convenience, abbreviate $\Tr_{q^2/q}(x)$ and $\Norm_{q^2/q}(y)$ by $\Tr(x)$ and $\Norm(y)$, respectively, for the rest of this subsection.

To determine the parameters and weight enumerator of $\cC_{f_1,g_1}$, we need the following results.

\begin{lemma}\label{lemma_num1}
For any $\alpha, \beta \in \F_q$, let $N^1_{\alpha,\beta} = \#\{x \in \F_{q^2} : \Tr(x)=\alpha, \Norm(x)= \beta\}$. Then for odd $q$,
$$
N^1_{\alpha,\beta} = 1- \eta(\alpha^2-4\beta),
$$
and for even $q$,
$$
N^1_{\alpha,\beta} =
\begin{cases}
1  & \text { if } \alpha=0, \\
1-(-1)^{\Tr_{q/2} \left(\frac{\beta}{\alpha^2}\right)} & \text { if } \alpha \neq 0.\\
\end{cases}
$$
\end{lemma}
\begin{IEEEproof}
By Vieta theorem, the equation set
$$\begin{cases}
\Tr(x) = x + x^q = \alpha, \\
\Norm(x) = x \cdot x^{q} = \beta
\end{cases}$$
has the same solutions as the equation $h(x)=x^2 - \alpha x +\beta =0$. Meanwhile, obviously $x$ and $x^q$ are two roots of $h(x) \in \F_q[x]$ over $\F_{q^2}$.
If $q$ is odd.

\begin{itemize}
\item When $\eta(\alpha^2-4\beta)=0$, from Lemma \ref{tr}, the root of $g(x)$ is $x=x^q = \alpha/2\in \F_q$. Then $N^1_{\alpha,\beta}=1$.
\item When $\eta(\alpha^2-4\beta)=1$, the roots of $g(x)$ are $x, x^q \in \F_q$ and $x\neq x^q$, a contradiction. Thus $N^1_{\alpha,\beta}=0$.
\item When $\eta(\alpha^2-4\beta)=-1$, the roots of $g(x)$ are $x, x^q \in \F_{q^2}$. Then $N^1_{\alpha,\beta}=2$.
\end{itemize}
Therefore, in this case, $N^1_{\alpha,\beta} = 1- \eta(\alpha^2-4\beta).$

If $q$ is even, the proof is almost the same by Lemma \ref{tr}, so the details are omitted.
\end{IEEEproof}

\begin{lemma}\label{lemma_num2}
For any $a \in \F_q$, $b \in \F_q^*$, let $N^2_{a,b} = \#\{x \in \F_{q^2} : a + \Tr(x) + b\Norm(x)=0\}$. Then $N^2_{a,b} \in \{1, q+1\}$, and $N^2_{a,b}=1$ if and only if $ab=1.$
\end{lemma}
\begin{IEEEproof}
By Lemma \ref{eta} and Lemma \ref{lemma_num1}, if $q$ is odd,
\[\begin{aligned}
N^2_{a,b}&=\sum\limits_{(\alpha,\beta) \in \F_q^2 \atop a+\alpha +b\beta=0} N^1_{\alpha,\beta}\\
&= \sum\limits_{(\alpha,\beta) \in \F_q^2 \atop a+\alpha +b\beta=0} \left(1- \eta(\alpha^2-4\beta)\right)\\
&= \sum\limits_{\beta \in \F_q } \left(1- \eta\left((a+b\beta)^2-4\beta\right)\right)\\
&= q- \sum\limits_{\beta \in \F_q } \eta\left(b^2 \beta^2 + (2ab -4)\beta +a^2\right)\\
&= \begin{cases}
q-(q-1)  & \text { if } (2ab-4)^2-4a^2b^2=0, \\
q-(-1)  & \text { if } (2ab-4)^2-4a^2b^2\neq 0, \\
\end{cases}\\
&= \begin{cases}
1  & \text { if } ab=1, \\
q+1  & \text { if } ab \neq 1.\\
\end{cases}
\end{aligned}\]
If $q$ is even,
\[\begin{aligned}
N^2_{a,b}&=\sum\limits_{(\alpha,\beta) \in \F_q^2 \atop a+\alpha +b\beta=0} N^1_{\alpha,\beta}\\
&= 1+ \sum\limits_{\alpha\in \F_q^{*} } \left(1-(-1)^{\Tr_{q/2} \left(\frac{a+ \alpha}{b\alpha^2}\right)}\right)\\
&= 1+ \sum\limits_{\alpha\in \F_q^{*} } \left(1-(-1)^{\Tr_{q/2} \left(\frac{a }{b\alpha^2}\right) + \Tr_{q/2} \left(\frac{1}{b\alpha}\right)}\right)\\
&= q-  \sum\limits_{\alpha\in \F_q^{*}} (-1)^{\Tr_{q/2}  \left(\frac{ab}{b^2\alpha^2}\right) + \Tr_{q/2} \left(\frac{1}{b^2\alpha^2}\right)}\\
&= q-  \sum\limits_{\alpha\in \F_q^{*}} (-1)^{\Tr_{q/2} \left(\frac{ab + 1}{b^2\alpha^2}\right)}\\
&= \begin{cases}
q-(q-1)=1  & \text { if } ab=1, \\
q-(-1)=q+1  & \text { if } ab \neq 1.\\
\end{cases}
\end{aligned}\]
The proof is now completed.
\end{IEEEproof}


Using the preparations above, we can determine the parameters and weight enumerator of ${\cal C}_{f_1,g_1}$.

\begin{theorem}\label{C_1}
Let notations be the same as before.
Then ${\cal C}_{f_1,g_1}$ is an optimal $[q^3+1, 3, q^3-q]$  code achieving the Griesmer bound over $\F_{q^2}$. Its weight enumerator is
$$1 + (q^2-1)(q^4-q^{3}+q^2)z^{q^3-q} + (q^2-1)(q^3+1)z^{q^3}.$$
Its dual ${\cal C}_{f_1,g_1}^{\perp}$ is a $[q^3+1, q^3-2, 3]$ almost MDS code.

\end{theorem}
\begin{IEEEproof}
For $m=2$, $f_1(x)=\Tr(x)$, $g_1(y)=\Norm(y)$, by the transitivity of trace functions, we have
$$\#D= q^3 + \frac{1}{q}\sum\limits_{z\in \F_q^*} \sum\limits_{x \in \F_{q^m}}\chi\left(z\Tr(x)\right) \sum\limits_{y \in \F_{q^m}}\chi\left(z\Norm(y)\right)= q^3.$$
Then from Lemma \ref{length}, the code length of ${\cal C}_{f_1,g_1}$ is $n=q^3+1.$

Let $S_1$ be the set of columns of matrix $G_{f_1,g_1}$.
We discuss the value of $\#({\cal H}_{\bf u}\cap S_1)$ for any line ${\cal H}_{\bf u}$ with nonzero ${\bf u}=(u_1,u_2,u_3)\in \F_{q^2}^3$ in the following cases.

\begin{itemize}
\item	If ${\bf u}=(u_1,0,0)$ for $u_1\in\F_{q^2}^*$. Obviously ${\cal H}_{\bf u}\cap S_1=\{(0,1,0)^{\top}\}$.

\item	If ${\bf u}=(0,u_2,0)$ for $u_2\in\F_{q^2}^*$. Then ${\cal H}_{\bf u}\cap S_1=\{(1,0,y)^{\top} : y \in \F_{q^2}, \Norm(y)=0\} = \{(1,0,0)^{\top}\}$.

\item	If ${\bf u}=(0,0,u_3)$ for $u_3\in\F_{q^2}^*$. Then ${\cal H}_{\bf u}\cap S_1= \{(0,1,0)^{\top}\} \cup \{(1,x,0)^{\top} : x \in \F_{q^2}, \Tr(x)=0\}$. Thus $\#({\cal H}_{\bf u}\cap S_1)=q+1$.

\item	If ${\bf u}=(u_1,u_2,0)$ for $u_1,u_2\in\F_{q^2}^*$. Then ${\cal H}_{\bf u}\cap S_1 = \{(1, -\frac{u_1}{u_2}, y)^{\top} : y \in \F_{q^2}, \Norm(y)= -\Tr(-\frac{u_1}{u_2})\}$. Thus $\#({\cal H}_{\bf u}\cap S_1)=1$ if $\Tr(-\frac{u_1}{u_2})=0$, and there are $(q^2-1)(q-1)$ choices for such $(u_1, u_2) \in (\F_{q^2}^*)^2$. $\#({\cal H}_{\bf u}\cap S_1)=q+1$ if $\Tr(-\frac{u_1}{u_2})\neq 0$, there are $(q^2-1)(q^2-q)$ choices for such $(u_1, u_2) \in (\F_{q^2}^*)^2$.

\item	If ${\bf u}=(u_1,0,u_3)$ for $u_1,u_3\in\F_{q^2}^*$. Then ${\cal H}_{\bf u}\cap S_1 = \{(0,1,0)^{\top}\} \cup \{(1,x, -\frac{u_1}{u_3})^{\top} : x \in \F_{q^2}, \Tr(x)= -\Norm(-\frac{u_1}{u_3})\}$. Thus $\#({\cal H}_{\bf u}\cap S_1)=q+1$.

\item	If ${\bf u}=(0,u_2,u_3)$ for $u_2,u_3\in\F_{q^2}^*$. Then ${\cal H}_{\bf u}\cap S_1 = \{(1,x, -\frac{u_2}{u_3}x)^{\top} : x \in \F_{q^2}, \Tr(x) + \Norm(-\frac{u_1}{u_3}x)=0\}$.
    By Lemma \ref{lemma_num2},
    $\#\{x \in \F_{q^2}, \Tr(x) + \Norm(-\frac{u_1}{u_3}x)=0\}
    =\#\{x \in \F_{q^2}, \Tr(x) + \Norm(-\frac{u_1}{u_3})\Norm(x)=0\}=q+1.$ Thus $\#({\cal H}_{\bf u}\cap S_1)=q+1$.

\item	If ${\bf u}=(u_1,u_2,u_3)$ for $u_1,u_2,u_3\in\F_{q^2}^*$. Then ${\cal H}_{\bf u}\cap S_1 = \{(1,-\frac{u_1}{u_2}-\frac{u_3}{u_2}y, y)^{\top} : y \in \F_{q^2}, \Tr(-\frac{u_1}{u_2}-\frac{u_3}{u_2}y) + \Norm(y)=0\}$.
    By Lemma \ref{lemma_num2},
    \[\begin{aligned}
    &\#\{y \in \F_{q^2}, \Tr(-\frac{u_1}{u_2}-\frac{u_3}{u_2}y) + \Norm(y)=0\}\\
    =&\#\{z \in \F_{q^2}, \Tr(-\frac{u_1}{u_2})+ \Tr(z) + \Norm(-\frac{u_2}{u_3}z)=0\}\\
    =&\#\{z \in \F_{q^2}, \Tr(-\frac{u_1}{u_2})+ \Tr(z) + \Norm(-\frac{u_2}{u_3})\Norm(z)=0\}\\
    =& \begin{cases}
       1 & \text { if } \Tr(-\frac{u_1}{u_2})\Norm(-\frac{u_2}{u_3})=1, \\
       q+1  & \text { if } \Tr(-\frac{u_1}{u_2})\Norm(-\frac{u_2}{u_3})\neq 1.\\
       \end{cases}
     \end{aligned}\]
     Thus, in this case, $\#({\cal H}_{\bf u}\cap S_1)\in \{1,q+1\}$ and $\#\{{\bf u}\in (\F_{q^2}^{*})^3 : \#({\cal H}_{\bf u}\cap S_1)=1\} = q(q^2-1)^2$.
\end{itemize}

For any nonzero $\textbf{u} \in \F_{q^2}^3$, by the foregoing discussions, we deduce that
$\#({\cal H}_{\bf u}\cap S_1) \in \{1,q+1\}$, and the number of ${\bf u}\in \F_{q^2}^3$ satisfying $\#({\cal H}_{\bf u}\cap S_1)=1$ is $\left(q^2-1)(1+1+q-1+q(q^2-1)\right)=(q^2-1)(q^3+1).$
The minimum distance and weight distribution of code ${\cal C}_{f_1,g_1}$ can be obtained directly from Lemma \ref{Weight-lem}.

Note that $\cC_{f_1,g_1}^{\perp}$ is of length $q^3 + 1$ and dimension $q^3-2$.
Since any two points of $S_1$ generate a line in PG$(2,q^2)$ and there exist three collinear points, such as $\{(0,1,0)^{\top},(1,0,0,)^{\top},(1,v,0)^{\top} \}$ for some $v \in \F_{q^2}^*$ with $\Tr(v)=0$. Hence, the minimum distance of $\cC_{f_1,g_1}^{\perp}$ is 3.
\end{IEEEproof}

\begin{remark}\label{remark2}
When $q=2$, $\cC_{f_1,g_1}$ is a quaternary Hermitian self-orthogonal code as every codeword of $\cC_{f_1,g_1}$ has weight divisible by two (\cite{Huffman}).

\end{remark}

Below, we present an example illustrating Theorem \ref{C_1}.

\begin{example}
\begin{itemize}
\item Let $q=2$. By \cite{Magma}, $\cC_{f_1,g_1}$ is an optimal $[9,3,6]$ code over $\F_4$ achieving the Griesmer bound with weight enumerator
$$1 + 36z^6 + 27z^8.$$
Its dual is a $[9, 6, 3]$ almost MDS code. These coincide with the results of Theorem \ref{C_1}.
\item Let $q=3$. By \cite{Magma}, $\cC_{f_1,g_1}$ is an optimal $[28,3,24]$ code over $\F_9$ achieving the Griesmer bound with weight enumerator
$$1 + 504z^{24} + 224z^{27}.$$
Its dual is an $[28, 25, 3]$ almost MDS code. These also coincide with the codes derived by Theorem \ref{C_1}.
\end{itemize}
\end{example}


The following theorem presents several specific information about the $q$-ary subfield code of ${\cal C}_{f_1,g_1}$, denoted by $\cC_{f_1,g_1}^{(q)}$, in terms of parameters, weight distribution and duality.

\begin{theorem}\label{sub_code1}
Let us keep the notation the same as before. Then,  the following results hold.
\begin{itemize}
  \item If $q$ is even, $\cC_{f_1,g_1}^{(q)}$ is a five-weight $q$-ary linear code with parameters $[q^3+1, 5, q^3-q^2-q ]$ and
the weight distribution is given in Table \ref{tableC1q-1}.
  \item If $q$ is odd, $\cC_{f_1,g_1}^{(q)}$ is a five-weight $q$-ary linear code with parameters $[q^3+1, 5, q^3-q^2-q+1]$ and
the weight distribution is given in Table \ref{tableC1q-2}.
  \item The dual $\cC_{f_1,g_1}^{(q)\bot}$ is always a $[q^3+1, q^3-4, 3]$ linear code.
\end{itemize}

\end{theorem}
\begin{IEEEproof}
The code length of $\cC_{f_1,g_1}^{(q)}$ is $q^3+1$ as shown in Theorem \ref{C_1}.
For any $a \in \F_q$, $(b,c) \in \F_{q^2}^2 \backslash (0,0)$, by $f_1(x)=\Tr(x)$, $g_1(y)=\Norm(y)$, we have
  \[\begin{aligned}
\Upsilon_{a,b,c}&=\sum\limits_{z\in \F_q^{*}} \chi(za) \sum\limits_{w \in \F_q^*}\sum\limits_{(x,y) \in \F_{q^2}^2}\chi\left(w \Tr(x) + w \Norm(y)\right)\chi'(zbx+zcy)\\
&=\sum\limits_{z\in \F_q^{*}} \chi(za) \sum\limits_{w \in \F_q^*} \sum\limits_{x \in \F_{q^2}}\chi'\left((w+zb)x\right) \sum\limits_{y \in \F_{q^2}}\chi\left(w\Norm(y)+\Tr(zcy)\right)\\
 &= \begin{cases}
       0  & \text { if } w\neq -zb, \\
        q^2\sum\limits_{y \in \F_{q^2}} \sum\limits_{z\in \F_q^{*}} \chi\left(z(a+\Tr(cy)-b\Norm(y))\right)  & \text { if } w = -zb,\\
       \end{cases}\\
  &= \begin{cases}
       0  & \text { if } b \in \F_{q^2} \backslash \F_q^*,\\
       q^3N_y-q^4    & \text { if }  b \in \F_q^*,\\
       \end{cases}
   \end{aligned}\]
 where $N_y = \#\{y \in \F_{q^2} : a+\Tr(cy)-b\Norm(y)=0\}$ and the last equality holds because for any $z \in \F_q^*$, there exists some $w \in \F_q^*$ such that $w=-zb$ if and only if $b \in \F_q^*$. From the properties of trace and norm functions,
 \begin{itemize}
 \item when $c=0$, $a=0$, $N_y=1;$
  \item when $c=0$, $a\neq 0$, $N_y=q+1;$
 \item when $c\neq0$, by Lemma \ref{lemma_num2},
  \[\begin{aligned}
  N_y &= \#\{y \in \F_{q^2} : a+\Tr(y)-b\Norm\left(\frac{1}{c}\right)\Norm(y)\}\\
  &= \begin{cases}
       1  & \text { if }~ ab\Norm\left(\frac{1}{c}\right)=-1, \\
       q+1  & \text { if }~ ab\Norm\left(\frac{1}{c}\right)\neq -1,\\
       \end{cases}\\
    &= \begin{cases}
       1  & \text { if }~  \Norm(c)=-ab, \\
       q+1  & \text { if }~ \Norm(c)\neq -ab.\\
       \end{cases}
   \end{aligned}\]
  \end{itemize}
  Thus,
  $$\Upsilon_{a,b,c}= \begin{cases}
       0  & \text { if } b \in \F_{q^2} \backslash \F_q^*,\\
       q^3-q^4  & \text { if }  b \in \F_q^*, \Norm(c)=-ab,\\
       q^3 & \text { if }  b \in \F_q^*, \Norm(c)\neq -ab,\\
       \end{cases}$$
where $a \in \F_q$, $(b,c) \in \F_{q^2}^2 \backslash (0,0)$.

Since $\#D = q^3$, by Lemma \ref{weight}, for any codeword $\textbf{c}_{a,b,c}$ in $\cC_{f_1,g_1}^{(q)}$ we deduce that
$$
  wt(\textbf{c}_{a,b,c})=\begin{cases}
0 & \text { if } a=b=c=0,\\
q^3  & \text { if } a\neq 0, b=c=0, \\
q^3-q^2  & \text { if } b=0, c \neq 0 ~\text{or}~b \in \F_{q^2} \backslash \F_q, \delta(b)=0,\\
q^3-q^2 +1  & \text { if } b \in \F_{q^2} \backslash \F_q, \delta(b)=1\\
q^3-q & \text { if }   b \in \F_q^*, \Norm(c)=-ab, \delta(b)=0,\\
q^3-q+1 & \text { if }   b \in \F_q^*, \Norm(c)=-ab, \delta(b)=1,\\
q^3-q^2-q  & \text { if }  b \in \F_q^*, \Norm(c)\neq -ab, \delta(b)=0,\\
q^3-q^2-q+1  & \text { if }  b \in \F_q^*, \Norm(c)\neq -ab, \delta(b)=1,\\
    \end{cases}
$$
  where $a \in \F_q$, $b,c \in \F_{q^2}.$

When $q$ is even, for any $b \in \F_{q^2}$, $\Tr(b)=0$ if and only if $b \in \F_q$. Then we further have
$$
  wt(\textbf{c}_{a,b,c})
= \begin{cases}
0,   & \text { with~ 1~ time}, \\
q^3-q^2-q  & \text { with~ $q^4-2q^3+q^2$~ time}, \\
q^3-q^2  & \text { with~$q^3-q$~ time},\\
q^3-q^2 +1 & \text { with~ $q^5-q^4$~ time}, \\
q^3-q  & \text { with~ $q^3-q^2$~ time}, \\
q^3  & \text { with~ $q-1$~ time}. \\
\end{cases}
$$

When $q$ is odd, for any $b \in \F_{q}$, $\Tr(b)=0$ if and only if $b=0$, meanwhile,
there are $q-1$ different $b$'s in $\F_{q^2} \backslash \F_q$ such that $\Tr(b)=0$. Then
$$
 wt(\textbf{c}_{a,b,c})= \begin{cases}
0,   & \text { with~ 1~ time}, \\
q^3-q^2-q+1  & \text { with~ $q^4-2q^3+q^2$~ time}, \\
q^3-q^2  & \text { with~$q^4-q$~ time},\\
q^3-q^2 +1 & \text { with~ $q^5-2q^4+q^3$~ time}, \\
q^3-q+1  & \text { with~ $q^3-q^2$~ time}, \\
q^3  & \text { with~ $q-1$~ time}. \\
\end{cases}
$$

The dimension of $\cC_{f_1,g_1}^{(q)}$ is 5 as $A_0=1.$

Note that $\cC_{f_1,g_1}^{(q)\perp}$ is of length $q^3 + 1$ and dimension $q^3-4$. It follows from Theorem \ref{C_1} and Lemma \ref{distance} that the minimal distance of $\cC_{f_1,g_1}^{(q)\perp}$ satisfies $d_1^{(q)\perp} \geq 3$. 
By the first four Pless power moments, one can derive that $d_1^{(q)\perp}  = 3$, whether $q$ is even or odd.
Then the proof is completed.
\end{IEEEproof}

\begin{table}[h!]
  \begin{center}
    \caption{The weight distribution of $\cC_{f_1,g_1}^{(q)}$ with $q$ even.}
    \begin{tabular}{c|c}
    \hline
      Weight & Multiplicity \\
      \hline
      0 & 1 \\
      $q^3-q^2-q$ & $q^4-2q^3+q^2$ \\
      $q^3-q^2$ & $q^3-q$ \\
      $q^3-q^2+1$ & $q^5-q^4$  \\
      $q^3-q$ & $q^3-q^2$  \\
      $q^3$ & $q-1$  \\
       \hline
    \end{tabular}
    \label{tableC1q-1}
  \end{center}
\end{table}

\begin{table}[h!]
  \begin{center}
    \caption{The weight distribution of $\cC_{f_1,g_1}^{(q)}$ with $q$ odd.}
    \begin{tabular}{c|c}
    \hline
      Weight & Multiplicity \\
      \hline
      0 & 1 \\
      $q^3-q^2-q+1$ & $q^4-2q^3+q^2$ \\
      $q^3-q^2$ & $q^4-q$ \\
      $q^3-q^2+1$ & $q^5-2q^4+q^3$  \\
      $q^3-q+1$ & $q^3-q^2$  \\
      $q^3$ & $q-1$  \\
       \hline
    \end{tabular}
    \label{tableC1q-2}
  \end{center}
\end{table}

\begin{example}\label{ex_sub_C1}
The following examples show that the subfield code $\cC_{f_1,g_1}^{(q)}$ has attractive properties. The optimality is obtained from the code tables at http://www.codetables.de/.
 \begin{itemize}
\item Let $q=2$. Then $\cC_{f_1,g_1}^{(q)}$ has parameters $[9,5,2]$, which is almost optimal. Its dual code $\cC_{f_1,g_1}^{(q)\perp}$ has parameters $[9, 4, 3]$, which is almost optimal.

\item Let $q=3$. Then $\cC_{f_1,g_1}^{(q)}$ has parameters $[28,5,16]$, which is almost optimal. Its dual code $\cC_{f_1,g_1}^{(q)\perp}$ has parameters $[28, 23, 3]$, which is optimal.

\item Let $q=4$. Then the dual code $\cC_{f_1,g_1}^{(q)\perp}$ has parameters $[65, 60, 3]$, which is optimal.

\item Let $q=5$. Then the dual code $\cC_{f_1,g_1}^{(q)\perp}$ has parameters $[126, 121, 3]$, which is optimal.
 \end{itemize}
\end{example}

From Lemma \ref{weight-2} and Theorem \ref{sub_code1}, we can directly get the following conclusions about the punctured code of the $q$-ary linear code $\cC_{f_1,g_1}^{(q)}$, which is denoted by $\bar{\cC}_{f_1,g_1}^{(q)}$.

\begin{theorem}\label{punc_code1}

The punctured code $\bar{\cC}_{f_1,g_1}^{(q)}$ is a four-weight $q$-ary linear code with parameters $[q^3, 5, q^3-q^2-q]$ and the weight enumerator is
$$1+ (q^4-2q^3+q^2)z^{q^3-q^2-q} + (q^5-q^4+q^3-q)z^{q^3-q^2} + (q^3-q^2)z^{q^3-q } + (q-1)z^{q^3}.$$
The dual $\bar{\cC}_{f_1,g_1}^{(q)\perp}$ is a $[q^3, q^3-5, 3]$ linear code for $q\geq 3$
and an $[8, 3, 4]$ binary code for $q=2$.
\end{theorem}

\begin{example} \label{ex_punc_C1}
The following examples show that the punctured code $\bar{\cC}_{f_1,g_1}^{(q)}$ is also attractive. The optimality are obtained from the code tables at http://www.codetables.de/.
 \begin{itemize}
\item Let $q=2$. Then $\bar{\cC}_{f_1,g_1}^{(q)}$ has parameters $[8,5,2]$, which is optimal. Its dual code $\bar{\cC}_{f_1,g_1}^{(q)\perp}$ has parameters $[8, 3, 4]$, which is optimal.

\item Let $q=3$. Then $\bar{\cC}_{f_1,g_1}^{(q)}$ has parameters $[27,5,15]$, which is almost optimal. Its dual code $\bar{\cC}_{f_1,g_1}^{(q)\perp}$ has parameters $[27, 22, 3]$, which is optimal.

\item Let $q=4$. Then the dual code $\bar{\cC}_{f_1,g_1}^{(q)\perp}$ has parameters $[64, 59, 3]$, which is optimal.

\item Let $q=5$. Then the dual code $\bar{\cC}_{f_1,g_1}^{(q)\perp}$ has parameters $[125, 120, 3]$, which is optimal.
 \end{itemize}
\end{example}

\begin{remark}
When $q=4$, $\bar{\cC}_{f_1,g_1}^{(q)}$ is a quaternary Hermitian self-orthogonal code as every codeword of $\cC_{f_1,g_1}$ has weight divisible by two (\cite{Huffman}).

\end{remark}

\subsection{$f_2(x)=\Tr_{q^m /q}(x)$ and $g_2(y)=\Tr_{q^m /q}(y^2)$}\label{Tr(y^2)}

In this subsection, let $m$ be a positive integer and $f_2(x)=\Tr_{q^m/q}(x)$, $g_2(y)=\Tr_{q^m/q}(y^2)$. Abbreviate $\Tr_{q^m/q}(x)$ by $\Tr(x)$ for the rest of this subsection.

In particular, when $m=1$, $f_2(x)=x$, $g_2(y)=y^2$. Then $\cC_{f_2,g_2}$ is a $[q+1,3,q-1]$ MDS code which has been studied in \cite{Heng2019}. So we focus on the case of $m \geq 2$ in this section.

We shall use the following lemma to obtain the parameters and weight enumerator of $\cC_{f_2,g_2}$.

\begin{lemma}\label{lemma_num3}
For any $u, v \in \F_{q^m}$, let $N^3_{u,v} = \#\{y \in \F_{q^m} : \Tr(y^2+uy+v)=0\}$. Then for $q=2^l$,
$$
 N^3_{u,v}
 = \begin{cases}
 2q^{m-1}  & \text { if } u \in \F_q^*, \Tr_{q^m/2}(\frac{v}{u^2})=0,\\
 0  & \text { if } u \in \F_q^*, \Tr_{q^m/2}(\frac{v}{u^2})=1,\\
 q^{m-1} & \text { if } u \in \F_{q^m} \backslash \F_q^*;
 \end{cases}
$$
and for $q=p^l$ with odd $p$, when $m$ is odd,
$$
 N^3_{u,v}=\begin{cases}
 q^{m-1} & \text { if } \Tr(v-\frac{u^2}{4})=0,\\
 q^{m-1} + q^{\frac{m-1}{2}} (-1)^{\frac{l(p-1)(m+1)}{4}}& \text { if } \Tr(v-\frac{u^2}{4}) \neq 0, \eta\left(\Tr(v-\frac{u^2}{4})\right)=1,\\
 q^m + q^{\frac{m-1}{2}} (-1)^{\frac{l(p-1)(m+1)+4}{4}}& \text { if } \Tr(v-\frac{u^2}{4}) \neq 0, \eta\left(\Tr(v-\frac{u^2}{4})\right)=-1;\\
 \end{cases}
$$
when $m$ is even,
$$
 N^3_{u,v}=\begin{cases}
 q^{m-1}+(q-1)q^{\frac{m-2}{2}}(-1)^{\frac{lm(p-1)+4}{4}}  & \text { if } \Tr(v-\frac{u^2}{4})=0,\\
 q^{m-1}+q^{\frac{m-2}{2}}(-1)^{\frac{lm(p-1)}{4}} & \text { if } \Tr(v-\frac{u^2}{4}) \neq 0.
 \end{cases}
$$

\end{lemma}
\begin{IEEEproof}
By the orthogonality relation of additive characters and the transitivity of trace functions, we have
 \[\begin{aligned}
 q N_{u,v}^3 &= \sum\limits_{y \in \F_{q^m}} \sum\limits_{z \in \F_q} \chi\left(z\left(\Tr(y^2+uy+v)\right)\right) \\
 &=q^m + \sum\limits_{z \in \F_q^*} \sum\limits_{y \in \F_{q^m}} \chi'(zy^2+zuy+zv).
 \end{aligned}\]

When $q$ is even, by Lemma \ref{character_even},
 \[\begin{aligned}
 q N_{u,v}^3
 &= \begin{cases}
 q^m + \chi'(\frac{v}{u^2})q^m  & \text { if } u \in \F_q^*, \\
 q^m & \text { if } u \in \F_{q^m} \backslash \F_q^*,
 \end{cases}\\
 &= \begin{cases}
 2q^m  & \text { if } u \in \F_q^*, \Tr_{q^m/2}(\frac{v}{u^2})=0,\\
 0  & \text { if } u \in \F_q^*, \Tr_{q^m/2}(\frac{v}{u^2})=1,\\
 q^m & \text { if } u \in \F_{q^m} \backslash \F_q^*.
 \end{cases}
 \end{aligned}\]

When $q$ is odd, by Lemma \ref{character_odd},
 \[\begin{aligned}
 q N_{u,v}^3
 &= q^m + \sum\limits_{z \in \F_q^*} \chi'\left(zv-z^2u^2(4z)^{-1}\right)\eta'(z) G(\eta', \chi')\\
 &= q^m + G(\eta', \chi')\sum\limits_{z \in \F_q^*} \chi'\left(z(v-\frac{u^2}{4})\right)\eta'(z).
 \end{aligned}\]

If $m$ is odd, we have $\eta'(z) = \eta(z)$ for $z \in \F_q^*$. Then by Lemma \ref{Gauss} we deduce that
 \[\begin{aligned}
 q N_{u,v}^3
 &= q^m + G(\eta', \chi')\sum\limits_{z \in \F_q^*} \chi\left(z\Tr(v-\frac{u^2}{4})\right)\eta(z)\\
 &= \begin{cases}
 q^m+G(\eta', \chi')\sum\limits_{z \in \F_q^*}\eta(z)  & \text { if } \Tr(v-\frac{u^2}{4})=0,\\
 q^m + G(\eta', \chi')G(\eta,\chi)\eta\left(\Tr(v-\frac{u^2}{4})\right)& \text { if } \Tr(v-\frac{u^2}{4}) \neq 0,
 \end{cases}\\
 &= \begin{cases}
 q^m & \text { if } \Tr(v-\frac{u^2}{4})=0,\\
 q^m + q^{\frac{m+1}{2}} (-1)^{\frac{l(p-1)(m+1)}{4}}& \text { if } \Tr(v-\frac{u^2}{4}) \neq 0, \eta\left(\Tr(v-\frac{u^2}{4})\right)=1,\\
 q^m + q^{\frac{m+1}{2}} (-1)^{\frac{l(p-1)(m+1)+4}{4}}& \text { if } \Tr(v-\frac{u^2}{4}) \neq 0, \eta\left(\Tr(v-\frac{u^2}{4})\right)=-1.\\
 \end{cases}
 \end{aligned}\]

 If $m$ is even, $\eta'(z)=1$ for any $z \in \F_q^*$. Then by Lemma \ref{Gauss} we deduce that
  \[\begin{aligned}
 q N_{u,v}^3
 &= q^m + G(\eta', \chi')\sum\limits_{z \in \F_q^*} \chi'\left(z(v-\frac{u^2}{4})\right)\\
 &= q^m + G(\eta', \chi')\sum\limits_{z \in \F_q^*} \chi\left(z\Tr(v-\frac{u^2}{4})\right)\\
 &= \begin{cases}
 q^m+(q-1)q^{\frac{m}{2}}(-1)^{\frac{lm(p-1)+4}{4}}  & \text { if } \Tr(v-\frac{u^2}{4})=0,\\
 q^m+q^{\frac{m}{2}}(-1)^{\frac{lm(p-1)}{4}} & \text { if } \Tr(v-\frac{u^2}{4}) \neq 0.
 \end{cases}
 \end{aligned}\]

The proof is now completed.
\end{IEEEproof}

We now settle the parameters and weight enumerator of the linear code ${\cal C}_{f_2,g_2}$ over $\F_{q^m}$.

\begin{theorem}\label{C_2}
Let $q=p^l$.
If $m \geq 2$, the following statements hold.
\begin{itemize}
\item When $q$ is even, ${\cal C}_{f_2,g_2}$ is a five-weight linear code over $\F_{q^m}$ with parameters $[q^{2m-1}+1, 3, q^{2m-1}-2q^{m-1}+1]$
 and the weight distribution is given in Table \ref{tableC2-1}.

\item When both $q$ and $m$ are odd, ${\cal C}_{f_2,g_2}$ is a five-weight linear code over $\F_{q^m}$ with parameters $[q^{2m-1}+1, 3, q^{2m-1}-q^{m-1}-q^{\frac{m-1}{2}}+1]$ and the weight distribution is given in Table \ref{tableC2-2}.

\item When $q$ is odd and $m$ is even, ${\cal C}_{f_2,g_2}$ is a four-weight linear code over $\F_{q^m}$ with parameters $[q^{2m-1}+1, 3]$ and weight distribution is given in Table \ref{tableC2-3}.

\item The dual ${\cal C}_{f_2,g_2}^{\perp}$ is always a $[q^{2m-1}+1, q^{2m-1}-2, 3]$ almost MDS code.
\end{itemize}

\end{theorem}
\begin{IEEEproof}
From $f_2(x)=\Tr(x)$, $g_2(y)=\Tr(y^2)$,
$$\#D= q^{2m-1} + \frac{1}{q}\sum\limits_{z\in \F_q^*} \sum\limits_{x \in \F_{q^m}}\chi(z\Tr(x)) \sum\limits_{y \in \F_{q^m}}\chi(z\Tr(y^2))= q^{2m-1}.$$
Then by Lemma \ref{length}, the code length of ${\cal C}_{f_2,g_2}$ is $n=q^{2m-1}+1$.

Like the proof of Theorem \ref{C_1}, let $S_2$ be the set of columns of the matrix $G_{f_2,g_2}$.
We discuss the value of $\#({\cal H}_{\bf u}\cap S_2)$ for any line ${\cal H}_{\bf u}$ with nonzero ${\bf u}\in \F_{q^m}^3$. Let $u_1,u_2,u_3 \in \F_{q^m}^*$, combined with Lemma \ref{lemma_num3}, we derive the following results.
\begin{itemize}

\item When $q$ is even,
$$\#({\cal H}_{\bf u}\cap S_2)=\begin{cases}
 0
 & \text { if } {\bf u}=(u_1,u_2,u_3), \frac{u_3}{u_2} \in \F_q^*, \Tr_{q^m/2}(\frac{u_1u_2}{u_3^2})=1,\\

 1
 & \text{ if } {\bf u}=(u_1,0,0),\\

 q^{m-1}
 & \text { if } {\bf u}=(0,u_2,0),\\
 & ~\text { or }{\bf u}=(u_1,u_2,0),\\
 & ~\text { or } {\bf u}=(0,u_2,u_3), \frac{u_3}{u_2} \in \F_{q^m}^* \backslash \F_q^*,\\
 & ~\text { or } {\bf u}=(u_1,u_2,u_3), \frac{u_3}{u_2} \in \F_{q^m}^* \backslash \F_q^*,\\

 q^{m-1}+1
 & \text { if } {\bf u}=(0,0,u_3),\\
 & ~\text { or } {\bf u}=(u_1,0,u_3),\\

 2q^{m-1}
 & \text { if } {\bf u}=(0,u_2,u_3), \frac{u_3}{u_2} \in \F_q^*,\\
 & ~\text { or } {\bf u}=(u_1,u_2,u_3), \frac{u_3}{u_2} \in \F_q^*,\Tr_{q^m/2}(\frac{u_1u_2}{u_3^2})=0.

 \end{cases}$$

Since
$$
\begin{aligned}
&\#\left\{(u_1,u_2,u_3) \in (\F_{q^m}^*)^3 : \frac{u_3}{u_2} \in \F_q^*,\Tr_{q^m/2}(\frac{u_1u_2}{u_3^2})=0 \right\}= (q^m-1)(q-1)(\frac{q^m}{2}-1).\\
&\#\left\{(u_1,u_2,u_3) \in (\F_{q^m}^*)^3 : \frac{u_3}{u_2} \in \F_q^*,\Tr_{q^m/2}(\frac{u_1u_2}{u_3^2})=1 \right\}= (q^m-1)(q-1)\frac{q^m}{2}.
\end{aligned}
$$
We deduce that
$$\#({\cal H}_{\bf u}\cap S_2)=\begin{cases}
 0
 & \text { with $(q^m-1)(q-1)\frac{q^m}{2}$ times},\\

 1
 & \text{ with $q^m-1$ times},\\

 q^{m-1}
 & \text { with $q^m(q^m-1)(q^m-q+1)$ times},\\

 q^{m-1}+1
 & \text { with $q^m(q^m-1)$ times},\\

 2q^{m-1}
 & \text { with $(q^m-1)(q-1)\frac{q^m}{2}$ times}.

 \end{cases}$$


\item When both $q$ and $m$ are odd,
$$
\#({\cal H}_{\bf u}\cap S_2)=\begin{cases}
1
& \text{ if } {\bf u}=(u_1,0,0),\\

q^{m-1}
& \text { if } {\bf u}=(0,u_2,0),\\
& ~\text { or }{\bf u}=(u_1,u_2,0), \Tr(-\frac{u_1}{u_2})=0,\\
& ~\text { or } {\bf u}=(0,u_2,u_3), \Tr(-\frac{u_3^2}{4u_2^2})=0,\\
& ~\text { or } {\bf u}=(u_1,u_2,u_3), \Tr(-\frac{u_1}{u_2}-\frac{u_3^2}{4u_2^2})=0,\\

q^{m-1}+1
& \text { if } {\bf u}=(0,0,u_3),\\
& ~\text { or } {\bf u}=(u_1,0,u_3),\\

q^{m-1} + q^{\frac{m-1}{2}} (-1)^{\frac{l(p-1)(m+1)}{4}}
& \text { if }{\bf u}=(u_1,u_2,0), \Tr(-\frac{u_1}{u_2}) \neq 0, \eta(\Tr(-\frac{u_1}{u_2}))=1,\\
&~\text { or } {\bf u}=(0,u_2,u_3), \Tr(-\frac{u_3^2}{4u_2^2}) \neq 0, \eta(\Tr(-\frac{u_3^2}{4u_2^2}))=1,\\
& ~\text { or } {\bf u}=(u_1,u_2,u_3), \Tr(-\frac{u_1}{u_2}-\frac{u_3^2}{4u_2^2}) \neq 0, \eta(\Tr(-\frac{u_1}{u_2}-\frac{u_3^2}{4u_2^2}))=1,\\

q^{m-1} + q^{\frac{m-1}{2}} (-1)^{\frac{l(p-1)(m+1)+4}{4}}
& \text { if }{\bf u}=(u_1,u_2,0), \Tr(-\frac{u_1}{u_2}) \neq 0, \eta(\Tr(-\frac{u_1}{u_2}))=-1,\\
&~\text { or } {\bf u}=(0,u_2,u_3), \Tr(-\frac{u_3^2}{4u_2^2}) \neq 0, \eta(\Tr(-\frac{u_3^2}{4u_2^2}))=-1,\\
&~\text { or } {\bf u}=(u_1,u_2,u_3),\Tr(-\frac{u_1}{u_2}-\frac{u_3^2}{4u_2^2}) \neq 0, \eta(\Tr(-\frac{u_1}{u_2}-\frac{u_3^2}{4u_2^2}))=-1.
 \end{cases}
$$
Since
$$
\begin{aligned}
&\#\left\{(u_1,u_2) \in (\F_{q^m}^*)^2 : \Tr(-\frac{u_1}{u_2})=0 \right\}=(q^m-1)(q^{m-1}-1).\\
&\#\left\{(u_2,u_3) \in (\F_{q^m}^*)^2 : \Tr(-\frac{u_3^2}{4u_2^2})=0 \right\}=(q^m-1)(q^{m-1}-1).\\
&\#\left\{(u_1,u_2,u_3) \in (\F_{q^m}^*)^3 : \Tr(-\frac{u_1}{u_2}-\frac{u_3^2}{4u_2^2})=0 \right\} =(q^m-1)(q^{2m-1}-2q^{m-1}+1).
\end{aligned}
$$
We deduce that
$$
\#({\cal H}_{\bf u}\cap S_2)=\begin{cases}
1
& \text{ with $q^m-1$ times,}\\

q^{m-1}
& \text { with $q^{2m-1}(q^m-1)$ times,}\\

q^{m-1}+1
& \text { with $q^m(q^m-1)$ times,}\\

q^{m-1} - q^{\frac{m-1}{2}}
& \text { with $\frac{q^{2m-1}(q^m-1)(q-1)}{2}$ times,}\\

q^{m-1} + q^{\frac{m-1}{2}} (-1)^{\frac{l(p-1)(m+1)+4}{4}}
& \text { with $\frac{q^{2m-1}(q^m-1)(q-1)}{2}$ times.}

 \end{cases}
$$

\item When $q$ is odd and $m$ is even,
$$
\begin{aligned}
 \#({\cal H}_{\bf u}\cap S_2)&=\begin{cases}
 1& \text{ if } {\bf u}=(u_1,0,0),\\

 q^{m-1}+1& \text { if } {\bf u}=(0,0,u_3),\\
          & ~\text { or } {\bf u}=(u_1,0,u_3),\\

 q^{m-1}+(q-1)q^{\frac{m-2}{2}}(-1)^{\frac{lm(p-1)+4}{4}}  & \text { if } {\bf u}=(0,u_2,0),\\
                        & ~\text { or } {\bf u}=(u_1,u_2,0), \Tr(-\frac{u_1}{u_2})=0,\\
                        & ~\text { or } {\bf u}=(0,u_2,u_3), \Tr(-\frac{u_3^2}{4u_2^2})=0,\\
                        & ~\text { or } {\bf u}=(u_1,u_2,u_3), \Tr(-\frac{u_1}{u_2}-\frac{u_3^2}{4u_2^2})=0,\\

 q^{m-1}+q^{\frac{m-2}{2}}(-1)^{\frac{lm(p-1)}{4}}
 & \text { if } {\bf u}=(u_1,u_2,0), \Tr(-\frac{u_1}{u_2})\neq 0,\\
 & ~\text { or } {\bf u}=(0,u_2,u_3), \Tr(-\frac{u_3^2}{4u_2^2})\neq 0,\\
 & ~\text { or } {\bf u}=(u_1,u_2,u_3), \Tr(-\frac{u_1}{u_2}-\frac{u_3^2}{4u_2^2}) \neq 0,
 \end{cases}\\
&= \begin{cases}
 1& \text{ with $q^m-1$ times,}\\

 q^{m-1}+1& \text { with $q^m(q^m-1)$ times,}\\

 q^{m-1}+(q-1)q^{\frac{m-2}{2}}(-1)^{\frac{lm(p-1)+4}{4}}  & \text { with $q^{2m-1}(q^m-1)$ times,}\\

 q^{m-1}+q^{\frac{m-2}{2}}(-1)^{\frac{lm(p-1)}{4}}
 & \text { with $(q^{2m}-q^{2m-1})(q^m-1)$ times.}
 \end{cases}
 \end{aligned}
$$

\end{itemize}

The minimum distance and weight distribution of code ${\cal C}_{f_2,g_2}$ can be obtained directly from Lemma \ref{Weight-lem}.

Note that $\cC_{f_2,g_2}^{\perp}$ is of length $q^{2m-1} + 1$ and dimension $q^{2m-1}-2$.
Since any two points of $S_2$ generate a line in PG$(2,q^m)$ and there exist three collinear points, such as $\{(0,1,0)^{\top},(1,0,0,)^{\top},(1,v,0)^{\top} \}$ for some $v \in \F_{q^m}^*$ with $\Tr(v)=0$. Thus the minimum distance of $\cC_{f_2,g_2}^{\perp}$ is 3.
\end{IEEEproof}

Now we give an example to illustrate Theorem \ref{C_2}.

\begin{example}
\begin{itemize}

\item Let $q=2$, $m=3$. By \cite{Magma}, $\cC_{f_2,g_2}$ is a $[33,3,25]$ linear code over $\F_{2^3}$ with weight enumerator
$$1 + 28z^{25} + 56z^{28} + 392z^{29} + 7z^{32} + 28z^{33}.$$
Its dual is a $[33, 30, 3]$ almost MDS code. It coincides with the results on the codes presented in Theorem \ref{C_2}.

\item  Let $q=3$, $m=3$. By \cite{Magma}, $\cC_{f_2,g_2}$ is a $[244, 3, 232]$ code over $\F_{3^3}$ with weight enumerator
$$1 + 6318z^{232} + 702z^{234} + 6318z^{235} + 6318z^{238} + 26z^{243}.$$
Its dual is a $[244, 241, 3]$ almost MDS code. It also coincides with the codes derived in Theorem \ref{C_2}.
\end{itemize}
\end{example}

\begin{table}[h!]
  \begin{center}
    \caption{The weight distribution of $\cC_{f_2,g_2}$ with $q$ even.}
    \begin{tabular}{c|c}
    \hline
      Weight & Multiplicity \\
      \hline
      0 & 1 \\
      $q^{2m-1}-2q^{m-1}+1$ & $\frac{q^m}{2}(q^m-1)(q-1)$ \\
      $q^{2m-1}-q^{m-1}$ & $q^m(q^m-1)$ \\
      $q^{2m-1}-q^{m-1}+1$ & $q^m(q^m-1)(q^m-q+1)$  \\
      $q^{2m-1}$ & $q^m-1$  \\
      $q^{2m-1}+1$ & $\frac{q^m}{2}(q^m-1)(q-1)$  \\
       \hline
    \end{tabular}
    \label{tableC2-1}
  \end{center}
\end{table}

\begin{table}[h!]
  \begin{center}
    \caption{The weight distribution of $\cC_{f_2,g_2}$ with $q$ odd and $m$ odd.}
    \begin{tabular}{c|c}
    \hline
      Weight & Multiplicity \\
      \hline
      0 & 1 \\
      $q^{2m-1}-q^{m-1}-q^{\frac{m-1}{2}}+1$ & $\frac{q^{2m-1}(q^m-1)(q-1)}{2}$ \\
      $q^{2m-1}-q^{m-1}$ & $q^m(q^m-1)$ \\
      $q^{2m-1}-q^{m-1}+1$ & $q^{2m-1}(q^m-1)$  \\
      $q^{2m-1}-q^{m-1}+q^{\frac{m-1}{2}}+1$ & $\frac{q^{2m-1}(q^m-1)(q-1)}{2}$  \\
      $q^{2m-1}$ & $q^m-1$  \\
       \hline
    \end{tabular}
    \label{tableC2-2}
  \end{center}
\end{table}

\begin{table}[h!]
  \begin{center}
    \caption{The weight distribution of $\cC_{f_2,g_2}$ with $q$ odd and $m$ even.}
    \begin{tabular}{c|c}
    \hline
      Weight & Multiplicity \\
      \hline
      0 & 1 \\
      $q^{2m-1}-q^{m-1} + (q-1)q^{\frac{m-2}{2}}(-1)^{\frac{lm(p-1)}{4}}+1$ & $q^{2m-1}(q^m-1)$ \\
      $q^{2m-1}-q^{m-1}+ q^{\frac{m-2}{2}}(-1)^{\frac{lm(p-1)+4}{4}}+1$ & $(q^{2m}-q^{2m-1})(q^m-1)$ \\
      $q^{2m-1}-q^{m-1}$ & $q^{m}(q^m-1)$  \\
      $q^{2m-1}$ & $q^m-1$  \\
       \hline
    \end{tabular}
    \label{tableC2-3}
  \end{center}
\end{table}

Define a class of exponential sums as $$\Omega(a,b,c)= \sum\limits_{z\in \F_q^{*}} \chi(za) \sum\limits_{y \in \F_{q^m}}\chi'(-zby^2+zcy), ~a \in \F_q, b, c \in \F_{q^m}.$$

\begin{lemma}\label{Omega}
For any $a \in \F_q, b \in \F_q^*$ and $c \in \F_{q^m}.$  If $q$ is even, we have
$$\Omega(a,b,c)=\begin{cases}
  q^m & \text { if } c \in \F_q^*, \Tr_{q/2}(\frac{ab}{c^2})=0, \\
 -q^m & \text { if } c \in \F_q^*, \Tr_{q/2}(\frac{ab}{c^2})=1, \\
 0 & \text { if } c \in \F_{q^m} \backslash \F_q^*.
 \end{cases}$$
 If both $q$ and $m$ are odd,
$$\Omega(a,b,c)= \begin{cases}
 0 & \text { if } a+\Tr(\frac{c^2}{4b})=0,\\
 q^{\frac{m+1}{2}} (-1)^{\frac{l(p-1)(m+1)}{4}}& \text { if } a+\Tr(\frac{c^2}{4b}) \neq 0, \eta'(-b)\eta\left(a+\Tr(\frac{c^2}{4b})\right)=1,\\
 q^{\frac{m+1}{2}} (-1)^{\frac{l(p-1)(m+1)+4}{4}}& \text { if } a+\Tr(\frac{c^2}{4b}) \neq 0, \eta'(-b)\eta\left(a+\Tr(\frac{c^2}{4b})\right)=-1.
 \end{cases}$$
 If $q$ is odd and $m$ is even,
 $$\Omega(a,b,c)=\begin{cases}
 (q-1)q^{\frac{m}{2}}(-1)^{\frac{lm(p-1)+4}{4}}  & \text { if } a+\Tr(\frac{c^2}{4b}) = 0,\\
 q^{\frac{m}{2}}(-1)^{\frac{lm(p-1)}{4}} & \text { if } a+\Tr(\frac{c^2}{4b}) \neq 0.\\
 \end{cases}$$

\end{lemma}
\begin{IEEEproof}
When $q$ is even, by Lemma \ref{character_even}, for any $z \in \F_q^{*}$
$$\sum\limits_{y \in \F_{q^m}}\chi'(zby^2+zcy)=
\begin{cases}
 q^m & \text { if } b=zc^2, \\
 0 & \text { otherwise.}
 \end{cases}$$
Then
 \[\begin{aligned}
 \Omega(a,b,c)&= \sum\limits_{z\in \F_q^{*}} \chi(za) \sum\limits_{y \in \F_{q^m}}\chi'(zby^2+zcy)\\
 &=\begin{cases}
 \chi(\frac{ab}{c^2}) q^m & \text { if } c \in \F_q^*, \\
 0 & \text { if } c \in \F_{q^m} \backslash \F_q^*,
 \end{cases}\\
 &=\begin{cases}
  q^m & \text { if } c \in \F_q^*, \Tr_{q/2}(\frac{ab}{c^2})=0, \\
 -q^m & \text { if } c \in \F_q^*, \Tr_{q/2}(\frac{ab}{c^2})=1, \\
 0 & \text { if } c \in \F_{q^m} \backslash \F_q^*,
 \end{cases}
 \end{aligned}\]
 where the second equality holds because there exists some $z \in \F_q^*$ such that $b=zc^2$ if and only if $c \in \F_q^*$.

When $q$ is odd, by Lemma \ref{character_odd},
 \[\begin{aligned}
 \Omega(a,b,c)&= \sum\limits_{z\in \F_q^{*}} \chi(za) \sum\limits_{y \in \F_{q^m}}\chi'(-zby^2+zcy)\\
 &= \sum\limits_{z\in \F_q^{*}} \chi(za) \chi'(-z^2c^2(-4zb)^{-1}) \eta'(-zb) G(\eta',\chi')\\
 &= G(\eta',\chi')\eta'(-b)\sum\limits_{z\in \F_q^{*}} \chi(za) \chi'(z\frac{c^2}{4b}) \eta'(z)\\
 &= G(\eta',\chi')\eta'(-b)\sum\limits_{z\in \F_q^{*}} \chi\left(za+z\Tr(\frac{c^2}{4b})\right) \eta'(z).
 \end{aligned}\]

If $m$ is odd,
 \[\begin{aligned}
  \Omega(a,b,c)
 &= \begin{cases}
 G(\eta', \chi')\eta'(-b)\sum\limits_{z \in \F_q^*}\eta(z)  & \text { if } a+\Tr(\frac{c^2}{4b})=0,\\
 G(\eta,\chi)G(\eta', \chi')\eta'(-b)\eta\left( a+\Tr(\frac{c^2}{4b}) \right)& \text { if } a+\Tr(\frac{c^2}{4b}) \neq 0,
 \end{cases}\\
 &= \begin{cases}
 0 & \text { if } a+\Tr(\frac{c^2}{4b})=0,\\
 q^{\frac{m+1}{2}} (-1)^{\frac{l(p-1)(m+1)}{4}}& \text { if } a+\Tr(\frac{c^2}{4b}) \neq 0, \eta'(-b)\eta\left(a+\Tr(\frac{c^2}{4b})\right)=1,\\
 q^{\frac{m+1}{2}} (-1)^{\frac{l(p-1)(m+1)+4}{4}}& \text { if } a+\Tr(\frac{c^2}{4b}) \neq 0, \eta'(-b)\eta\left(a+\Tr(\frac{c^2}{4b})\right)=-1.\\
 \end{cases}
 \end{aligned}\]

 If $m$ is even, $\eta'(-b)=1$ for any $b \in \F_q^*$, then
  \[\begin{aligned}
  \Omega(a,b,c)
 &=G(\eta', \chi') \sum\limits_{z \in \F_q^*} \chi\left(z(a+\Tr(\frac{c^2}{4b}))\right)\\
 &= \begin{cases}
 (q-1)q^{\frac{m}{2}}(-1)^{\frac{lm(p-1)+4}{4}}  & \text { if } a+\Tr(\frac{c^2}{4b})=0,\\
 q^{\frac{m}{2}}(-1)^{\frac{lm(p-1)}{4}} & \text { if } a+\Tr(\frac{c^2}{4b})\neq 0.
 \end{cases}
 \end{aligned}\]
The proof is now completed.

\end{IEEEproof}



We now determine the parameters of the $q$-ary subfield code of ${\cal C}_{f_2,g_2}$.

\begin{theorem}\label{sub_code2}
Let $q=p^l$.
If $m \geq 2$, the following statements hold.
\begin{itemize}
\item When both $q$ and $m$ are even.
\begin{itemize}
\item If $q=2$,
${\cal C}_{f_2,g_2}^{(q)}$ is a three-weight binary linear code with parameters $[2^{2m-1}+1, 2m, 2^{2m-2}]$ and
weight distribution in Table \ref{tableC2q-1}.
 The dual ${\cal C}_{f_2,g_2}^{(q)\perp}$ is an $[2^{2m-1}+1, 2^{2m-1}-2m+1, 3]$ binary code.

 \item If $q=2^l$, $l\geq 2$, ${\cal C}_{f_2,g_2}^{(q)}$ is a four-weight $q$-ary linear code with parameters $[q^{2m-1}+1, 2m+1, (q-2)q^{2m-2}]$
 and the weight distribution in Table \ref{tableC2q-2}.
  The dual ${\cal C}_{f_2,g_2}^{(q)\perp}$ is an $[q^{2m-1}+1, q^{2m-1}-2m, 3]$ linear code.
\end{itemize}

\item When $q$ is even and $m$ is odd.
\begin{itemize}
\item ${\cal C}_{f_2,g_2}^{(q)}$ is always a five-weight $q$-ary linear code with parameters $[q^{2m-1}+1, 2m+1, (q-2)q^{2m-2}+1]$
 and the weight distribution in Table \ref{tableC2q-3}.

\item If $q=2$, the dual ${\cal C}_{f_2,g_2}^{(q)\perp}$ is an optimal $[2^{2m-1}+1, 2^{2m-1}-2m, 4]$ binary code with respect to the sphere-packing bound.

 \item If $q=2^l$, $l\geq 2$, the dual ${\cal C}_{f_2,g_2}^{(q)\perp}$ is an $[q^{2m-1}+1, q^{2m-1}-2m, 3]$ linear code over $\F_{q}$.
\end{itemize}

\item When both $q$ and $m$ are odd.
\begin{itemize}
\item If $p \mid m$,
${\cal C}_{f_2,g_2}^{(q)}$ is a five-weight $q$-ary linear code with parameters $[q^{2m-1}+1, 2m+1, (q-1)q^{2m-2}-q^{\frac{3m-3}{2}}]$
 and the weight distribution in Table \ref{tableC2q-4}.

 \item If $p \nmid m$,
${\cal C}_{f_2,g_2}^{(q)}$ is a five-weight $q$-ary linear code with parameters $[q^{2m-1}+1, 2m+1, (q-1)q^{2m-2}-q^{\frac{3m-3}{2}}+1]$
 and the weight distribution in Table \ref{tableC2q-5}.

  \item The dual ${\cal C}_{f_2,g_2}^{(q)\perp}$ is always a $[q^{2m-1}+1, q^{2m-1}-2m, 3]$ linear code.
 \end{itemize}

\item When $q$ is odd and $m$ is even.
\begin{itemize}
\item If $p \mid m$,
${\cal C}_2^{(q)}$ is a five-weight $q$-ary linear code with parameters $[q^{2m-1}+1, 2m+1]$ and the weight distribution in Table \ref{tableC2q-6}.

 \item If $p \nmid m$,
${\cal C}_2^{(q)}$ is a five-weight $q$-ary linear code with parameters $[q^{2m-1}+1, 2m+1]$ and the weight distribution in Table \ref{tableC2q-7}.

 \item The dual ${\cal C}_2^{(q)\perp}$ is always an $[q^{2m-1}+1, q^{2m-1}-2m, 3]$ linear code over $\F_{q}$.
 \end{itemize}

\end{itemize}

\end{theorem}
\begin{IEEEproof}
The code length of $\cC_{f_2,g_2}^{(q)}$ is $q^{2m-1}+1$ as show in Theorem \ref{C_2}.
For any $a \in \F_q$, $(b,c) \in \F_{q^m}^2 \backslash (0,0)$, by $f_2(x)=\Tr(x)$, $g_2(y)=\Tr(y^2)$, we have
  \[\begin{aligned}
\Upsilon_{a,b,c}&=\sum\limits_{z\in \F_q^{*}} \chi(za) \sum\limits_{w \in \F_q^*}\sum\limits_{(x,y) \in \F_{q^m}^2}\chi(w \Tr(x) + w \Tr(y^2))\chi'(zbx+zcy)\\
&=\sum\limits_{z\in \F_q^{*}} \chi(za) \sum\limits_{w \in \F_q^*} \sum\limits_{x \in \F_{q^m}}\chi'\left((w+zb)x\right) \sum\limits_{y \in \F_{q^m}}\chi'(wy^2+zcy)\\
 &= \begin{cases}
 0  & \text { if } w\neq -zb, \\
 q^m\sum\limits_{z\in \F_q^{*}} \chi(za) \sum\limits_{y \in \F_{q^m}}\chi'(-zby^2+zcy)  & \text { if } w =-zb,\\
       \end{cases}\\
  &= \begin{cases}
       0  & \text { if } b \in \F_{q^m} \backslash \F_q^*,\\
       q^m \Omega(a,b,c)   & \text { if }  b \in \F_q^*,\\
       \end{cases}
   \end{aligned}\]
where $\Omega(a,b,c)$ is shown in Lemma \ref{Omega}.

Since $\#D = q^{2m-1}$, by Lemma \ref{weight}, for any codeword $\textbf{c}_{a,b,c}$ in $\cC_{f_2,g_2}^{(q)}$,
\begin{itemize}
\item when $q$ is even, we deduce that
$$
  wt(\textbf{c}_{a,b,c})= \begin{cases}
  0        & \text { if }   a=b=c=0,\\

  q^{2m-1} & \text { if } a \neq 0,b=0,c=0,\\
            & ~\text { or }  b \in \F_q^*, c \in \F_q^*, \Tr_{q/2}(\frac{ab}{c^2})=1, \delta(b)=0,\\

  q^{2m-1}+1 & \text { if }  b \in \F_q^*, c \in \F_q^*, \Tr_{q/2}(\frac{ab}{c^2})=1, \delta(b)=1,\\

  (q-1)q^{2m-2} & \text { if } b \in \F_{q^m}\backslash\F_q, \delta(b)=0, \\
                 & ~\text{ or } b=0, c\neq 0,\\
                 & ~\text{ or } b \in \F_q^*, c \in \F_{q^m} \backslash \F_q^*, \delta(b)=0,\\

 (q-1)q^{2m-2} +1 & \text { if } b \in \F_{q^m}\backslash\F_q, \delta(b)=1, \\
                 & ~\text{ or } b \in \F_q^*, c \in \F_{q^m} \backslash \F_q^*, \delta(b)=1,\\

 (q-2)q^{2m-2}  & \text{ if } b \in \F_q^*, c \in \F_q^*, \Tr_{q/2}(\frac{ab}{c^2})=0, \delta(b)=0,\\

 (q-2)q^{2m-2}+1  & \text{ if } b \in \F_q^*, c \in \F_q^*, \Tr_{q/2}(\frac{ab}{c^2})=0, \delta(b)=1,\\

 \end{cases}
 $$
where $a \in \F_q$, $b,c \in \F_{q^m}.$

\begin{itemize}
\item If $2 \mid m$, for any $b \in \F_q^*$, $\delta(b)=0$. Then
$$wt(\textbf{c}_{a,b,c})= \begin{cases}
0  & \text { with~ $1$~ time}, \\
 q^{2m-1}   & \text { with~ $(q-1)(q^2-q+2)/2$ time}, \\
(q-1)q^{2m-2}  & \text { with~ $q^{2m}-q^3+2q^2-2q$~ time}, \\
(q-1)q^{2m-2}+1  & \text { with~$q^{2m+1}-q^{2m}$~ time},\\
(q-2)q^{2m-2} & \text { with~ $q(q-1)^2/2$~ time}.
\end{cases}
$$
\item If $2 \nmid m$, for any $b \in \F_q^*$, $\delta(b)=1$. Then
$$wt(\textbf{c}_{a,b,c})= \begin{cases}
 0  & \text { with~ $1$~ time}, \\
 q^{2m-1}   & \text { with~ $q-1$ time}, \\
 q^{2m-1}+1  & \text { with~ $q(q-1)^2/2$ time}, \\
(q-1)q^{2m-2}  & \text { with~ $q^{2m}-q$~ time}, \\
(q-1)q^{2m-2}+1  & \text { with~$q^{2m+1}-q^{2m}-q^3+2q^2-q$~ time},\\
(q-2)q^{2m-2}+1 & \text { with~ $q(q-1)^2/2$~ time}. \\

\end{cases}
$$
\end{itemize}

\item When both $q$ and $m$ are odd, we deduce that
$$
 wt(\textbf{c}_{a,b,c})= \begin{cases}
  0 & \text { if } a=b=c=0,\\

  q^{2m-1} & \text { if } a\neq 0, b=c=0,\\

 (q-1)q^{2m-2} & \text { if }  b \in \F_{q^m} \backslash \F_q, \delta(b)=0,\\
               & ~\text{ or } b=0, c \neq 0,\\
               & ~\text{ or } b \in \F_q^*, a+\Tr(\frac{c^2}{4b})=0, \delta(b)=0,\\

 (q-1)q^{2m-2}+1 & \text { if }  b \in \F_{q^m} \backslash \F_q, \delta(b)=1,\\
                 & ~\text{ or } b \in \F_q^*, a+\Tr(\frac{c^2}{4b})=0, \delta(b)=1,\\

 (q-1)q^{2m-2} + q^{\frac{3m-3}{2}} (-1)^{\frac{l(p-1)(m+1)+4}{4}}& \text { if } b \in \F_q^*, a+\Tr(\frac{c^2}{4b}) \neq 0, \\& ~~~~\eta'(-b)\eta\left(a+\Tr(\frac{c^2}{4b})\right)=1, \delta(b)=0,\\

  (q-1)q^{2m-2} + q^{\frac{3m-3}{2}} (-1)^{\frac{l(p-1)(m+1)+4}{4}}+1& \text { if } b \in \F_q^*, a+\Tr(\frac{c^2}{4b}) \neq 0, \\& ~~~~\eta'(-b)\eta\left(a+\Tr(\frac{c^2}{4b})\right)=1, \delta(b)=1,\\

 (q-1)q^{2m-2} + q^{\frac{3m-3}{2}} (-1)^{\frac{l(p-1)(m+1)}{4}}& \text { if } b \in \F_q^*, a+\Tr(\frac{c^2}{4b}) \neq 0, \\& ~~~~\eta'(-b)\eta\left(a+\Tr(\frac{c^2}{4b})\right)=-1, \delta(b)=0,\\

  (q-1)q^{2m-2} + q^{\frac{3m-3}{2}} (-1)^{\frac{l(p-1)(m+1)}{4}}+1& \text { if } b \in \F_q^*, a+\Tr(\frac{c^2}{4b}) \neq 0, \\& ~~~~\eta'(-b)\eta\left(a+\Tr(\frac{c^2}{4b})\right)=-1, \delta(b)=1,\\
 \end{cases}
$$
where $a \in \F_q$, $b,c \in \F_{q^m}.$

\begin{itemize}
\item  If $p\mid m$,
$$
 wt(\textbf{c}_{a,b,c})=  \begin{cases}
    0        & \text { with~ 1 time},\\
   q^{2m-1} & \text { with~ $q-1$ time},\\
  (q-1)q^{2m-2} & \text { with~ $q^{2m}-q^{m+2}+2q^{m+1}-q^m-q$ time},\\
 (q-1)q^{2m-2}+1 & \text { with~ $q^{2m+1}-q^{2m}$ time},\\
 (q-1)q^{2m-2} + q^{\frac{3m-3}{2}} (-1)^{\frac{l(p-1)(m+1)+4}{4}}& \text { with~ $(q^{m+2}-2q^{m+1}+q^m)/2$ time},\\
 (q-1)q^{2m-2} + q^{\frac{3m-3}{2}} (-1)^{\frac{l(p-1)(m+1)}{4}}& \text { with~ $(q^{m+2}-2q^{m+1}+q^m)/2$ time}.
 \end{cases}
 $$

\item If $p\nmid m$,
$$ wt(\textbf{c}_{a,b,c})= \begin{cases}
   0        & \text { with~ 1 time},\\
   q^{2m-1} & \text { with~ $q-1$ time},\\
  (q-1)q^{2m-2} & \text { with~ $q^{2m}-q$ time},\\
 (q-1)q^{2m-2}+1 & \text { with~ $q^{2m+1}-q^{2m}-q^{m+2}+2q^{m+1}-q^m$ time},\\
 (q-1)q^{2m-2} + q^{\frac{3m-3}{2}} (-1)^{\frac{l(p-1)(m+1)+4}{4}}+1& \text { with~ $(q^{m+2}-2q^{m+1}+q^m)/2$ time},\\
 (q-1)q^{2m-2} + q^{\frac{3m-3}{2}} (-1)^{\frac{l(p-1)(m+1)}{4}}+1& \text { with~ $(q^{m+2}-2q^{m+1}+q^m)/2$ time}.
 \end{cases}
$$
\end{itemize}

\item When $q$ is odd and $m$ is even, we deduce that
$$
wt(\textbf{c}_{a,b,c})= \begin{cases}
 0        & \text { if } a = 0, b=c=0,\\

 q^{2m-1} & \text { if } a\neq0, b=c=0,\\

 (q-1)q^{2m-2} & \text { if }  b \in \F_{q^m} \backslash \F_q, \delta(b)=0, \\
               & \text{ or } b=0, c \neq 0, \\

  (q-1)q^{2m-2}+1 & \text { if }  b \in \F_{q^m} \backslash \F_q, \delta(b)=1, \\

(q-1)q^{2m-2}+ (q-1)q^{\frac{3m-4}{2}}(-1)^{\frac{lm(p-1)}{4}}  & \text { if }  b \in \F_q^*,a+\Tr(\frac{c^2}{4b}) =0, \delta(b)=0,\\

(q-1)q^{2m-2}+ (q-1)q^{\frac{3m-4}{2}}(-1)^{\frac{lm(p-1)}{4}}+1 & \text { if }  b \in \F_q^*,a+\Tr(\frac{c^2}{4b}) =0, \delta(b)=1,\\

(q-1)q^{2m-2}+q^{\frac{3m-4}{2}}(-1)^{\frac{lm(p-1)+4}{4}} & \text { if }  b \in \F_q^*,a+\Tr(\frac{c^2}{4b})\neq 0,  \delta(b)=0,\\

(q-1)q^{2m-2}+q^{\frac{3m-4}{2}}(-1)^{\frac{lm(p-1)+4}{4}}+1 & \text { if }  b \in \F_q^*,a+\Tr(\frac{c^2}{4b})\neq 0,  \delta(b)=1,\\

 \end{cases}
$$
where $a \in \F_q$, $b,c \in \F_{q^m}.$

\begin{itemize}
\item If $p\mid m$,
$$
wt(\textbf{c}_{a,b,c})= \begin{cases}
   0        & \text { with~ 1 time},\\
   q^{2m-1} & \text { with~ $q-1$ time},\\
 (q-1)q^{2m-2} & \text { with~ $q^{2m}-q^{m+2}+q^{m+1}-q$ time},\\

 (q-1)q^{2m-2}+1 & \text { with~ $q^{2m+1}-q^{2m}$ time}, \\
 (q-1)q^{2m-2}+ (q-1)q^{\frac{3m-4}{2}}(-1)^{\frac{lm(p-1)}{4}} & \text{ with~ $q^{m+1}-q^m$ time},\\
 (q-1)q^{2m-2}+q^{\frac{3m-4}{2}}(-1)^{\frac{lm(p-1)+4}{4}} & \text { with~ $q^{m}(q-1)^2$ time}.\\
       \end{cases}
$$

\item If $p\nmid m$,
$$
wt(\textbf{c}_{a,b,c})= \begin{cases}
   0        & \text { with~ 1 time},\\
   q^{2m-1} & \text { with~ $q-1$ time},\\
 (q-1)q^{2m-2} & \text { with~ $q^{2m}-q$ time},\\
 (q-1)q^{2m-2}+1 & \text { with~ $q^{2m+1}-q^{2m}-q^{m+2}+q^{m+1}$ time}, \\
 (q-1)q^{2m-2}+ (q-1)q^{\frac{3m-4}{2}}(-1)^{\frac{lm(p-1)}{4}}+1 & \text{ with~ $q^{m+1}-q^m$ time},\\
 (q-1)q^{2m-2}+q^{\frac{3m-4}{2}}(-1)^{\frac{lm(p-1)+4}{4}}+1 & \text { with~ $q^{m}(q-1)^2$ time}.\\
       \end{cases}
$$
\end{itemize}

\end{itemize}

When $q=2$ and $m$ is even, observe that $(q-2)q^{2m-2}=0$, the codeword with Hamming weight 0 occurs two times if $(a, b, c)$ runs through $\F_2 \times \F_{2^m} \times \F_{2^m}$. Thus, in this subcase, every codeword in $C_{f_2,g_2}^{(q)}$ repeats two times, the dimension of $\cC_2^{(2)}$ is $2m$. In other subcases, the dimension of $\cC_{f_2,g_2}^{(q)}$ is $2m+1$ as $A_0=1.$

 The code length and dimension of $\cC_{f_2,g_2}^{(q)\perp}$ are obvious. In addition, it follows from Theorem \ref{C_2} and Lemma \ref{distance} that the minimal distance of $\cC_{f_2,g_2}^{(q)\perp}$ satisfies $d_2^{(q)\perp} \geq 3$. 
 \begin{itemize}
\item When $q=2$ and $m$ is odd, $\cC_{f_2,g_2}^{(q)\perp}$ is of length $2^{2m-1}+1$ and dimension $2^{2m-1}-2m$. By the sphere-packing bound of codes, we have
    $$2^{2^{2m-1}+1} \geq 2^{2^{2m-1}-2m} \left(\sum\limits_{i=0}^{\lfloor d_2^{(2)\perp}-1 \rfloor} \binom{2^{2m-1}+1}{i} \right).$$
    Consequently,
    $$2^{2m+1} \geq  \left(\sum\limits_{i=0}^{\lfloor d_2^{(2)\perp}-1 \rfloor} \binom{2^{2m-1}+1}{i} \right).$$
    Then it's easy to verify that $d_2^{(q)\perp} \leq 4$.
    By the first four Pless power moments, one can derive that $d_2^{(q)\perp} =4$.
\item For the other subcases, by the first four Pless power moments, we can always derive that $d_2^{(q)\perp} =3$.
 \end{itemize}
 The proof is now completed.
\end{IEEEproof}

\begin{table}[h!]
  \begin{center}
    \caption{The weight distribution of $\cC_{f_2,g_2}^{(q)}$ with $q=2$ and $m$ even.}
    \begin{tabular}{c|c}
    \hline
      Weight & Multiplicity \\
      \hline
      0 & 1 \\
      $2^{2m-2}$ & $2^{2m-1}-2$ \\
      $2^{2m-2}+1$ & $2^{2m}-2^{2m-1}$ \\
      $2^{2m-1}$ & 1  \\
       \hline
    \end{tabular}
    \label{tableC2q-1}
  \end{center}
\end{table}

\begin{table}[h!]
  \begin{center}
    \caption{The weight distribution of $\cC_{f_2,g_2}^{(q)}$ with $q=2^l$, $l\geq2$ and $m$ even.}
    \begin{tabular}{c|c}
    \hline
      Weight & Multiplicity \\
      \hline
      0 & 1 \\
      $(q-2)q^{2m-2}$ & $q(q-1)^2/2$ \\
      $(q-1)q^{2m-2}$ & $q^{2m}-q^3+2q^2-2q$ \\
      $(q-1)q^{2m-2}+1$ & $q^{2m+1}-q^{2m}$ \\
      $q^{2m-1}$ & $(q-1)(q^2-q+2)/2$  \\
       \hline
    \end{tabular}
    \label{tableC2q-2}
  \end{center}
\end{table}

\begin{table}[h!]
  \begin{center}
    \caption{The weight distribution of $\cC_{f_2,g_2}^{(q)}$ with $q$ even and $m$ odd.}
    \begin{tabular}{c|c}
    \hline
      Weight & Multiplicity \\
      \hline
      0 & 1 \\
      $(q-2)q^{2m-2}+1$ & $q(q-1)^2/2$ \\
      $(q-1)q^{2m-2}$ & $q^{2m}-q$ \\
      $(q-1)q^{2m-2}+1$ & $q^{2m+1}-q^{2m}-q^3+2q^2-q$\\
      $q^{2m-1}$ & $q-1$ \\
      $q^{2m-1}+1$ & $q(q-1)^2/2$\\
       \hline
    \end{tabular}
    \label{tableC2q-3}
  \end{center}
\end{table}

\begin{table}[h!]
  \begin{center}
    \caption{The weight distribution of $\cC_{f_2,g_2}^{(q)}$ with $q$ odd, $m$ odd and $p \mid m$.}
    \begin{tabular}{c|c}
    \hline
      Weight & Multiplicity \\
      \hline
      0 & 1 \\
      $(q-1)q^{2m-2} - q^{\frac{3m-3}{2}}$ & $(q^{m+2}-2q^{m+1}+q^m)/2$ \\
      $(q-1)q^{2m-2}$ & $q^{2m}-q^{m+2}+2q^{m+1}-q^m-q$ \\
      $(q-1)q^{2m-2}+1$ & $q^{2m+1}-q^{2m}$\\
      $(q-1)q^{2m-2} + q^{\frac{3m-3}{2}}$ & $(q^{m+2}-2q^{m+1}+q^m)/2$ \\
      $q^{2m-1}$ & $q-1$\\
       \hline
    \end{tabular}
    \label{tableC2q-4}
  \end{center}
\end{table}

\begin{table}[h!]
  \begin{center}
    \caption{The weight distribution of $\cC_{f_2,g_2}^{(q)}$ with $q$ odd, $m$ odd and $p \nmid m$.}
    \begin{tabular}{c|c}
    \hline
      Weight & Multiplicity \\
      \hline
      0 & 1 \\
      $(q-1)q^{2m-2} - q^{\frac{3m-3}{2}}+1$ & $(q^{m+2}-2q^{m+1}+q^m)/2$ \\
      $(q-1)q^{2m-2}$ & $q^{2m}-q$ \\
      $(q-1)q^{2m-2}+1$ & $q^{2m+1}-q^{2m}-q^{m+2}+2q^{m+1}-q^m$\\
      $(q-1)q^{2m-2} + q^{\frac{3m-3}{2}}+1$ & $(q^{m+2}-2q^{m+1}+q^m)/2$ \\
      $q^{2m-1}$ & $q-1$\\
       \hline
    \end{tabular}
    \label{tableC2q-5}
  \end{center}
\end{table}

\begin{table}[h!]
  \begin{center}
    \caption{The weight distribution of $\cC_{f_2,g_2}^{(q)}$ with $q$ odd, $m$ even and $p \mid m$.}
    \begin{tabular}{c|c}
    \hline
      Weight & Multiplicity \\
      \hline
      0 & 1 \\
      $(q-1)q^{2m-2} + q^{\frac{3m-4}{2}}(-1)^{\frac{lm(p-1)+4}{4}}$ & $q^m(q-1)^2$ \\
      $(q-1)q^{2m-2} + (q-1)q^{\frac{3m-4}{2}}(-1)^{\frac{lm(p-1)}{4}}$ & $q^{m+1}-q^m$ \\
      $(q-1)q^{2m-2}$ & $q^{2m}-q^{m+2}+ q^{m+1}-q$ \\
      $(q-1)q^{2m-2}+1$ & $q^{2m+1}-q^{2m}$\\
      $q^{2m-1}$ & $q-1$\\
       \hline
    \end{tabular}
    \label{tableC2q-6}
  \end{center}
\end{table}

\begin{table}[h!]
  \begin{center}
    \caption{The weight distribution of $\cC_{f_2,g_2}^{(q)}$ with $q$ odd, $m$ even and $p \nmid m$.}
    \begin{tabular}{c|c}
    \hline
      Weight & Multiplicity \\
      \hline
      0 & 1 \\
      $(q-1)q^{2m-2} + q^{\frac{3m-4}{2}}(-1)^{\frac{lm(p-1)+4}{4}}+1$ & $q^m(q-1)^2$ \\
      $(q-1)q^{2m-2} + (q-1)q^{\frac{3m-4}{2}}(-1)^{\frac{lm(p-1)}{4}}+1$ & $q^{m+1}-q^m$ \\
      $(q-1)q^{2m-2}$ & $q^{2m}-q$ \\
      $(q-1)q^{2m-2}+1$ & $q^{2m+1}-q^{2m}-q^{m+2}+ q^{m+1}$\\
      $q^{2m-1}$ & $q-1$\\
       \hline
    \end{tabular}
    \label{tableC2q-7}
  \end{center}
\end{table}

\begin{example}\label{ex_sub_C2}
The following examples show that the subfield code $\cC_{f_2,g_2}^{(q)}$ has attractive properties. Tables indicate the optimality at http://www.codetables.de/.
 \begin{itemize}
\item Let $q=2$, $m=2$. Then $\cC_{f_2,g_2}^{(q)}$ has parameters $[9,4,4]$, which is optimal. Its dual code $\cC_{f_2,g_2}^{(q)\perp}$ has parameters $[9, 5, 3]$, which is almost optimal.

\item Let $q=2$, $m=3$. Then the dual code $\cC_{f_2,g_2}^{(q)\perp}$ has parameters $[33,26,4]$, which is optimal.

\item Let $q=2$, $m=4$. Then $\cC_{f_2,g_2}^{(q)}$ has parameters $[129,8,64]$, which is optimal. Its dual code $\cC_{f_2,g_2}^{(q)\perp}$ has parameters $[129,121,3]$, which is optimal.

\item Let $q=4$, $m=2$. Then the dual code $\cC_{f_2,g_2}^{(q)\perp}$ has parameters $[65, 60, 3]$, which is optimal.

\item Let $q=3$, $m=2$. Then the dual code $\cC_{f_2,g_2}^{(q)\perp}$ has parameters $[28,23,3]$, which is optimal.

\item Let $q=5$, $m=2$. Then the dual code $\cC_{f_2,g_2}^{(q)\perp}$ has parameters $[126,121,3]$, which is optimal.
 \end{itemize}
\end{example}


We can directly get the following results from Lemma \ref{weight-2} and Theorem \ref{sub_code2}.

\begin{theorem}\label{punc_code2}
Let $q=p^l$.
If $m \geq 2$, the following statements hold.
\begin{itemize}
\item When $q$ is even.
\begin{itemize}
\item If $q=2$, $\bar{\cC}_{f_2,g_2}^{(2)}$ is an optimal $[2^{2m-1}, 2m, 2^{2m-2}]$ binary code achieving the Griesmer bound with
weight enumerator
$$1+ (2^{2m}-2)z^{2^{2m-2}} +z^{2^{2m-1}}.$$
 The dual $\bar{\cC}_{f_2,g_2}^{(2)\perp}$ is an optimal $[2^{2m-1}, 2^{2m-1}-2m, 4]$ binary code with respect to the sphere-packing bound.

 \item If $q=2^l$, $l\geq 2$, $\bar{\cC}_{f_2,g_2}^{(q)}$ is a three-weight $q$-ary linear code with parameters $[q^{2m-1}, 2m+1, (q-2)q^{2m-2}]$
 and the weight distribution in Table \ref{tableC2q-8}.
  The dual $\bar{\cC}_{f_2,g_2}^{(q)\perp}$ is a $[q^{2m-1}, q^{2m-1}-2m-1, 3]$ linear code over $\F_{q}$.

\end{itemize}

\item When both $q$ and $m$ are odd. $\bar{\cC}_{f_2,g_2}^{(q)}$ is a four-weight $q$-ary linear code with parameters $[q^{2m-1}, 2m+1, (q-1)q^{2m-2}-q^{\frac{3m-3}{2}}]$
 and the weight distribution in Table \ref{tableC2q-9}.
  The dual $\bar{\cC}_{f_2,g_2}^{(q)\perp}$ is a $[q^{2m-1}, q^{2m-1}-2m-1, 3]$ linear code over $\F_{q}$.

\item When $q$ is odd and $m$ is even. $\bar{\cC}_{f_2,g_2}^{(q)}$ is a four-weight $q$-ary linear code with parameters $[q^{2m-1}, 2m+1]$ and the weight distribution in Table \ref{tableC2q-10}.
  The dual $\bar{\cC}_{f_2,g_2}^{(q)\perp}$ is also a $[q^{2m-1}, q^{2m-1}-2m-1, 3]$ linear code over $\F_{q}$.

\end{itemize}
\end{theorem}

Reed-Muller codes are classical error-correcting codes which have also been extensively studied and used in many areas related to coding theory. We shall only consider binary Reed-Muller codes $\mathcal{RM}(r,n)$,  which can be viewed as the set of all binary vectors of length $2^n$ associated with multivariate binary polynomials $f(x_1,\dots,x_n)$ of algebraic degree at most $r$ (see, e.g. \cite{Sloane77})  and whose dual codes $\mathcal{RM}(r,n)^\perp$ are $\mathcal{RM}(n-r-1,n)$.

\begin{remark}\label{re4}
Note that, when $q=2$, $\bar{\cC}_{f_2,g_2}^{(2)}$ is actually the first order Reed-Muller code $\mathcal{RM}(1,2m-1)$. When $q=4$, $m \geq 2$, by Table \ref{tableC2q-8} every codeword of $\bar{\cC}_{f_2,g_2}^{(4)}$ has weight divisible by two, therefore, it is a $[2^{4m-2}, 2m+1, 2^{4m-3}]$ quaternary Hermitian self-orthogonal code (\cite{Huffman}).

\end{remark}

\begin{table}[h!]
  \begin{center}
    \caption{The weight distribution of $\bar{\cC}_{f_2,g_2}^{(q)}$ with $q=2^l$, $l\geq2$.}
    \begin{tabular}{c|c}
    \hline
      Weight & Multiplicity \\
      \hline
      0 & 1 \\
      $(q-2)q^{2m-2} $ & $q(q-1)^2/2$ \\
      $(q-1)q^{2m-2}$ & $q^{2m+1}-q^3+2q^2-2q$ \\
      $q^{2m-1}$ & $(q-1)(q^2-q-2)/2$\\
       \hline
    \end{tabular}
    \label{tableC2q-8}
  \end{center}
\end{table}

\begin{table}[h!]
  \begin{center}
    \caption{The weight distribution of $\bar{\cC}_{f_2,g_2}^{(q)}$ with $q$ odd and $m$ odd.}
    \begin{tabular}{c|c}
    \hline
      Weight & Multiplicity \\
      \hline
      0 & 1 \\
      $(q-1)q^{2m-2} -q^{\frac{3m-3}{2}}$ & $q^m(q-1)^2/2$ \\
      $(q-1)q^{2m-2}$ & $q^{2m+1}-q^{m+2}+2q^{m+1}-q^m-q$ \\
      $(q-1)q^{2m-2} +q^{\frac{3m-3}{2}}$ & $q^m(q-1)^2/2$ \\
      $q^{2m-1}$ & $q-1$\\
       \hline
    \end{tabular}
    \label{tableC2q-9}
  \end{center}
\end{table}

\begin{table}[h!]
  \begin{center}
    \caption{The weight distribution of $\bar{\cC}_{f_2,g_2}^{(q)}$ with $q$ odd and $m$ even.}
    \begin{tabular}{c|c}
    \hline
      Weight & Multiplicity \\
      \hline
      0 & 1 \\
      $(q-1)q^{2m-2} +q^{\frac{3m-4}{2}}(-1)^{\frac{lm(p-1)+4}{4}}$ & $q^{m+2}-2q^{m+1}+q^m$ \\
      $(q-1)q^{2m-2}+(q-1)q^{\frac{3m-4}{2}}(-1)^{\frac{lm(p-1)}{4}}$ & $q^{m+1}-q^m$ \\
      $(q-1)q^{2m-2}$ & $q^{2m+1}-q^{2m}+q^{m+1}-q$ \\
      $q^{2m-1}$ & $q-1$\\
       \hline
    \end{tabular}
    \label{tableC2q-10}
  \end{center}
\end{table}

\begin{example} \label{ex_punc_C2}
The following examples show that the subfield code $\bar{\cC}_{f_2,g_2}^{(q)}$ has attractive specificities. The code tables at http://www.codetables.de/ claim its optimality.
 \begin{itemize}
\item Let $q=2$, $m=2$. Then $\bar{\cC}_{f_2,g_2}^{(q)}$ has parameters $[8,4,4]$, which is optimal. Its dual code $\bar{\cC}_{f_2,g_2}^{(q)\perp}$ has parameters $[8, 4, 4]$, which is  optimal.

\item Let $q=2$, $m=3$. Then $\bar{\cC}_{f_2,g_2}^{(q)}$ has parameters $[32,6,16]$, which is optimal. Its dual code $\bar{\cC}_{f_2,g_2}^{(q)\perp}$ has parameters $[32,26, 4]$, which is  optimal.

\item Let $q=2$, $m=4$. Then $\bar{\cC}_{f_2,g_2}^{(q)}$ has parameters $[128,8,64]$, which is optimal. Its dual code $\bar{\cC}_{f_2,g_2}^{(q)\perp}$ has parameters $[128,120,4]$, which is optimal.

\item Let $q=4$, $m=2$. Then the dual code $\bar{\cC}_{f_2,g_2}^{(q)\perp}$ has parameters $[64,59,3]$, which is optimal.

\item Let $q=3$, $m=3$. Then the dual code $\bar{\cC}_{f_2,g_2}^{(q)\perp}$ has parameters $[243,236,3]$, which is optimal.

\item Let $q=3$, $m=2$. Then the dual code $\bar{\cC}_{f_2,g_2}^{(q)\perp}$ has parameters $[27,22,3]$, which is optimal.

\item Let $q=5$, $m=2$. Then the dual code $\bar{\cC}_{f_2,g_2}^{(q)\perp}$ has parameters $[125,120,3]$, which is optimal.
 \end{itemize}
\end{example}

\subsection{$f_3(x)=\Tr_{2^m/2}(x)$ and $g_3(y)=\Tr_{2^m/2}\left(A(y)\right)$ with $A(y)$ an almost bent function}\label{AB1}

In this section, let $q=2$, $m$ be odd and $f_3(x)=\Tr_{2^m/2}(x)$, $g_3(y)=\Tr_{2^m/2}(A(y))$, where $A(y)$ is an almost bent function from $\F_{2^m}$ to itself. For convenience, we abbreviate $\Tr_{2^m/2}(x)$ by $\Tr(x)$ for the rest of this subsection.

\begin{theorem}\label{sub_code3}
The code $\cC_{f_3,g_3}^{(2)}$ is a five-weight binary code with parameters $[2^{2m-1}+1, 2m+1, 2^{2m-2}- 2^{\frac{3m-3}{2}}+1]$ and
the weight distribution is given in Table \ref{tableC3q-1}.
Its  dual $\cC_{f_3,g_3}^{(2)\bot}$ is is an optimal $[2^{2m-1}+1, 2m, 4]$
binary code with respect to the sphere-packing bound.
\end{theorem}

\begin{IEEEproof}
For $f_3(x)=\Tr(x)$, $g_3(y)=\Tr(A(y))$, by the transitivity of trace functions, we have
$$\#D= 2^{2m-1} + \frac{1}{2} \sum\limits_{x \in \F_{2^m}}\chi(\Tr(x)) \sum\limits_{y \in \F_{2^m}}\chi(\Tr(A(y)))= 2^{2m-1}.$$
Then from Lemma \ref{length}, the code length of ${\cal C}_{f_3,g_3}^{(2)}$ is $n=2^{2m-1}+1.$

For any $a \in \F_2$, $(b,c) \in \F_{2^m}^2 \backslash (0,0)$,
  \[\begin{aligned}
\Upsilon_{a,b,c}&=\sum\limits_{z\in \F_2^{*}} \chi(za) \sum\limits_{w \in \F_2^*}\sum\limits_{(x,y) \in \F_{2^m}^2}\chi(w \Tr(x) + w \Tr(A(y)))\chi'(zbx+zcy)\\
&=(-1)^a \sum\limits_{x \in \F_{2^m}}\chi'\left((1+b)x\right) \sum\limits_{y \in \F_{2^m}}\chi'(A(y)+cy)\\
 &= \begin{cases}
       0  & \text { if } b\neq 1, \\
       (-1)^a2^m W_A(1,c)  & \text { if } b = 1,\\
       \end{cases}
   \end{aligned}\]
 where $W_A(1,c)= 0$ or $\pm 2^{\frac{m+1}{2}}$ from the definition of almost Bent function.

Since $m$ is odd, $\Tr(1)=m \neq 0$. By Lemma \ref{weight}, for any codeword $\textbf{c}_{a,b,c}$ in $\cC_{f_3,g_3}^{(2)}$ we deduce that
  \[\begin{aligned}
  wt(\textbf{c}_{a,b,c})&=\begin{cases}
0 & \text { if } a=b=c=0,\\
2^{2m-1}  & \text { if } a\neq 0, b=c=0, \\
2^{2m-2}  & \text { if } b=0, c\neq0,\\
           & ~\text { or } b \notin \{0,1\}, \delta(b)=0, \\

2^{2m-2}+1  & \text { if } b\neq1,\delta(b)=1,\\
           & ~\text { or } b=1, W_A(1,c)=0\\
2^{2m-2}- 2^{\frac{3m-3}{2}}+1 & \text { if }   b=1, a=0,  W_A(1,c)=2^{\frac{m+1}{2}},\\
                              & ~\text { or }   b=1, a=1,  W_A(1,c)=-2^{\frac{m+1}{2}},\\
2^{2m-2}+ 2^{\frac{3m-3}{2}}+1 & \text { if }   b=1, a=0,  W_A(1,c)=-2^{\frac{m+1}{2}},\\
                              &~ \text { or }   b=1, a=1,  W_A(1,c)=2^{\frac{m+1}{2}},\\
    \end{cases}\\
&=\begin{cases}
0 & \text { with 1 time,}\\
2^{2m-1}  & \text { with 1 time,}\\
2^{2m-2}  & \text { with $2^{2m}-2$ times,} \\
2^{2m-2}+1  & \text { with $i_1$ times,}\\
2^{2m-2}- 2^{\frac{3m-3}{2}}+1 & \text {  with $i_2$ times,}\\
2^{2m-2}+ 2^{\frac{3m-3}{2}}+1 & \text {  with $i_3$ times.}\\
    \end{cases}
 \end{aligned}\]
It's obvious that $i_2=i_3$ and the dimension of $\cC_{f_3,g_3}^{(2)}$ is $2m+1$ as $A_0=1.$

Since any two columns of matrix $G_{f_3,g_3}$ are linearly independent. It follows from Lemma \ref{distance} that the minimal distance $d_3^{(2)\perp}$ of $\cC_{f_3,g_3}^{(2)\perp}$ satisfies $d_3^{(2)\perp} \geq 3$. 
Consequently, from the first three Pless power moments, one can derive that $i_1=2^{2m}-2^m$ and $i_2=i_3= 2^{m-1}$.

Note that $\cC_{f_3,g_3}^{(2)\perp}$ is of length $q^{2m-1} + 1$ and dimension $q^{2m-1}-2m$.
By the sphere-packing bound of codes, we have
    $$2^{2^{2m-1}+1} \geq 2^{2^{2m-1}-2m} \left(\sum\limits_{i=0}^{\lfloor d_3^{(2)\perp}-1 \rfloor} \binom{2^{2m-1}+1}{i} \right).$$
    Consequently,
    $$2^{2m+1} \geq  \left(\sum\limits_{i=0}^{\lfloor d_3^{(2)\perp}-1 \rfloor} \binom{2^{2m-1}+1}{i} \right).$$
    Then it's easy to verify that $d_3^{(2)\perp} \leq 4$.
    By the first four Pless power moments, one can derive that $d_3^{(2)\perp} =4$.
\end{IEEEproof}

\begin{table}[h!]
  \begin{center}
    \caption{The weight distribution of $\cC_{f_3,g_3}^{(2)}$.}
    \begin{tabular}{c|c}
    \hline
      Weight & Multiplicity \\
      \hline
      0 & 1 \\
      $2^{2m-2}-2^{\frac{3m-3}{2}}+1$ & $2^{m-1}$ \\
      $2^{2m-2}$ & $2^{2m}-2$ \\
      $2^{2m-2}+1$ & $2^{2m}-2^m$ \\
      $2^{2m-2}+2^{\frac{3m-3}{2}}+1$ & $2^{m-1}$  \\
      $2^{2m-1}$ & 1 \\
       \hline
    \end{tabular}
    \label{tableC3q-1}
  \end{center}
\end{table}

We can directly get the following results from Lemma \ref{weight-2} and Theorem \ref{sub_code3}.

\begin{theorem}\label{punc_code3}
The linear code $\bar{{\cal C}}_{f_3,g_3}^{(2)}$ is a four-weight binary linear code with parameters $[2^{2m-1}, 2m+1, 2^{2m-2}-2^{\frac{3m-3}{2}}]$ and the weight distribution in Table \ref{tableC3q-2}.
The dual code $\bar{{\cal C}}_{f_3,g_3}^{(2)\perp}$ is an optimal $[2^{2m-1}, 2^{2m-1}-2m-1, 4]$ binary code with respect to the sphere-packing bound.
\end{theorem}

\begin{table}[h!]
  \begin{center}
    \caption{The weight distribution of $\bar{{\cal C}}_{f_3,g_3}^{(2)}$.}
    \begin{tabular}{c|c}
    \hline
      Weight & Multiplicity \\
      \hline
      0 & 1 \\
      $2^{2m-2}-2^{\frac{3m-3}{2}}$ & $2^{m-1}$ \\
      $2^{2m-2}$ & $2^{2m+1}-2^m-2$ \\
      $2^{2m-2}+2^{\frac{3m-3}{2}}$ & $2^{m-1}$  \\
      $2^{2m-1}$ & 1 \\
       \hline
    \end{tabular}
   \label{tableC3q-2}
  \end{center}
\end{table}

\subsection{$f_4(x)=\Tr_{2^m/2}(A_1(x))$ and $g_4(y)=\Tr_{2^m/2}(A_2(y))$ with $A_1$, $A_2$ are two different almost bent functions}\label{AB2}

In this section, let $q=2$, $m$ be odd and $f_4(x)=\Tr_{2^m/2}(A_1(x))$, $g_4(y)=\Tr_{2^m/2}(A_2(y))$, where $A_1$, $A_2$ are two different almost bent functions from $\F_{2^m}$ to itself.
We mainly focus on the parameters and the weight distribution of the punctured code $\bar{{\cal C}}_{f_4,g_4}^{(2)}$.
For convenience, we abbreviate $\Tr_{2^m/2}(x)$ by $\Tr(x)$ for the rest of this subsection.



\begin{theorem}\label{puncode4}
The punctured code $\bar{{\cal C}}_{f_4,g_4}^{(2)}$ is a four-weight binary code with parameters $[2^{2m-1}+\frac{W}{2}, 2m+1, 2^{2m-2}+\frac{W}{4}-2^{m-1}],$ where $W\in \{0, -2^{m+1}, 2^{m+1}\}$.
The weight distribution is shown in Table \ref{tableC4q-1}.
The dual $\bar{{\cal C}}_{f_4,g_4}^{(2)\perp}$ is an optimal $[2^{2m-1}+\frac{W}{2}, 2^{2m-1}+\frac{W}{2}-2m-1, 4]$ binary code with respect to the sphere-packing bound.
\end{theorem}
\begin{IEEEproof}
The code length of $\bar{{\cal C}}_{f_4,g_4}^{(2)}$ with $f_4(x)=\Tr(A_1(x))$, $g_4(y)=\Tr(A_2(y))$, is
$$n=\#D= 2^{2m-1} + \frac{1}{2} \sum\limits_{x \in \F_{2^m}}\chi(\Tr(A_1(x))) \sum\limits_{y \in \F_{2^m}}\chi(\Tr(A_2(y)))= 2^{2m-1}+ \frac{W}{2},$$
where $W = W_{A_1}(1,0)W_{A_2}(1,0)=0$ or $\pm 2^{m+1}$.

For any $a \in \F_2$, $(b,c) \in \F_{2^m}^2 \backslash (0,0)$,
  \[\begin{aligned}
\Upsilon_{a,b,c}&=\sum\limits_{z\in \F_2^{*}} \chi(za) \sum\limits_{w \in \F_2^*}\sum\limits_{(x,y) \in \F_{2^m}^2}\chi(w \Tr(A_1(x)) + w \Tr(A_2(y)))\chi'(zbx+zcy)\\
&=(-1)^a \sum\limits_{x \in \F_{2^m}}\chi'\left(A_1(x)+bx\right) \sum\limits_{y \in \F_{2^m}}\chi'(A_2(y)+cy)\\
 &= (-1)^a W',
   \end{aligned}\]
 where $W'=W_{A_1}(1,b)W_{A_2}(1,c)= 0$ or $\pm 2^{m+1}$.

 By Lemma \ref{weight-2}, for any codeword $\bar{\textbf{c}}_{a,b,c}$ in $\bar{\cC}_{f_4,g_4}^{(2)}$ we deduce that
  \[\begin{aligned}
  wt(\bar{\textbf{c}}_{a,b,c})&=\begin{cases}
0 & \text { if } a=b=c=0,\\
2^{2m-1}+ \frac{W}{2}  & \text { if } a\neq 0, b=c=0, \\
2^{2m-2}+ \frac{W}{4}  & \text { if $b$ and $c$ are not all 0, } W'=0,\\

2^{2m-2}+ \frac{W}{4} + 2^{m-1}  & \text { if $b$ and $c$ are not all 0, } a=0, ~ W'=-2^{m+1},\\
           & ~~~~~~~~~~~~~~~~~~~~~~~~~\text { or } a=1, ~W'=2^{m+1},\\
2^{2m-2}+ \frac{W}{4} - 2^{m-1}  & \text { if $b$ and $c$ are not all 0, } a=0, ~ W'=2^{m+1},\\
           & ~~~~~~~~~~~~~~~~~~~~~~~~~\text { or } a=1,~ W'=-2^{m+1},\\
    \end{cases}\\
&=\begin{cases}
0 & \text { with 1 time,}\\
2^{2m-1} + \frac{W}{2}  & \text { with 1 time,}\\
2^{2m-2}+ \frac{W}{4}  & \text { with $i_1$ times,}\\
2^{2m-2}+ \frac{W}{4} + 2^{m-1} & \text {  with $i_2$ times,}\\
2^{2m-2}+ \frac{W}{4} - 2^{m-1} & \text {  with $i_3$ times.}\\
    \end{cases}
 \end{aligned}\]
It's obvious that $i_2=i_3$ and the dimension of $\bar{\cC}_{f_4,g_4}^{(2)}$ is $2m+1$ as $A_0=1.$

Let $\bar{G}_{f_4,g_4}$ be the submatrix obtained by deleting the first column of $G_{f_4,g_4}$.
Since any two columns of matrix $\bar{G}_{f_4,g_4}$ are linear independent. It follows from Lemma \ref{distance} that the minimal distance of $\bar{\cC}_{f_4,g_4}^{(2)\perp}$ satisfies $\bar{d}_4^{(2)\perp} \geq 3$. 
Then by the first three Pless power moments, one can derive that $i_1=3\cdot2^{2m-1}-2+\frac{W^2}{2^{2m+1}}$ and $i_2=i_3= 2^{2m-2}-\frac{W^2}{2^{2m+2}}$.

Note that $\bar{\cC}_{f_4,g_4}^{(2)\perp}$ is of length $2^{2m-1} + \frac{W}{2}$ and dimension $2^{2m-1} + \frac{W}{2}-2m-1$.
By the sphere-packing bound of codes, we have
    $$2^{2^{2m-1} + \frac{W}{2}} \geq 2^{2^{2m-1} + \frac{W}{2}-2m-1} \left(\sum\limits_{i=0}^{\lfloor \bar{d}_4^{(2)\perp}-1 \rfloor} \binom{2^{2m-1} + \frac{W}{2}}{i} \right).$$
    Consequently,
    $$2^{2m+1} \geq  \left(\sum\limits_{i=0}^{\lfloor \bar{d}_4^{(2)\perp}-1 \rfloor} \binom{2^{2m-1} + \frac{W}{2}}{i} \right).$$
    It's easy to verify that $\bar{d}_4^{(2)\perp} \leq 4$.
    By the first four Pless power moments, one can derive that $\bar{d}_4^{(2)\perp} =4$.
\end{IEEEproof}

\begin{table}[h!]
  \begin{center}
    \caption{The weight distribution of $\bar{{\cal C}}_{f_4,g_4}^{(2)}$.}
    \begin{tabular}{c|c}
    \hline
      Weight & Multiplicity \\
      \hline
      0 & 1 \\
      $2^{2m-2}+\frac{W}{4}-2^{m-1}$ & $2^{2m-2}-\frac{W^2}{2^{2m+2}}$ \\
      $2^{2m-2}+\frac{W}{4}$ & $3\cdot2^{2m-1}-2+\frac{W^2}{2^{2m+1}}$ \\
      $2^{2m-2}+\frac{W}{4}+2^{m-1}$ & $2^{2m-2}-\frac{W^2}{2^{2m+2}}$  \\
      $2^{2m-1}+\frac{W}{2}$ & 1 \\
       \hline
    \end{tabular}
   \label{tableC4q-1}
  \end{center}
\end{table}

The following is a list of known almost bent monomials $A(x)=x^t$ on $\F_{2^m}$ for an odd $m$:
 \begin{itemize}
\item $t = 2^r + 1$, where $\gcd(r, m) = 1$ (\cite{Gold1968});

\item $t = 2^{2r} - 2^r + 1$, where $r \geq 2$ and $\gcd(m, h) = 1$ (\cite{Kasami1971});

\item $t=2^{\frac{m-1}{2}}+3$, where $m$ is odd \cite{Kasami1971};

\item $t= 2^{\frac{m-1}{2}} + 2^{\frac{m-1}{4}}-1$, where $m~\equiv~1 ~(\text{mod } 4)$ (\cite{Hollmann2001,Hou2004});

\item $t= 2^{\frac{m-1}{2}} + 2^{\frac{3m-1}{4}}-1$, where $m~\equiv~3 ~(\text{mod } 4)$ (\cite{Hollmann2001,Hou2004}).
\end{itemize}

All almost bent monomials $A(x) = x^t$ for $t$ in the list above are permutation polynomials on $\F_{2^m}$. Hence, the length of $\bar{{\cal C}}_{f_4,g_4}^{(2)}$ is $n=2^{2m-1}$ when at least one of $A_i(x)$, $i=1,2$, is such a monomial. We obtain the following results by substituting the value of $n$ into Theorem \ref{puncode4}.

\begin{corollary}\label{punc_code4}
Let $f_4(x)=\Tr(A_1(x))$, $g_4(y)=\Tr(A_2(y))$, where $A_i(x)$, $i=1,2$ be distinct almost bent functions, and at least one of them is a monomial $x^t$ for some integer $t$ in the list above. Then
$\bar{{\cal C}}_{f_4,g_4}^{(2)}$ is a four-weight binary code with parameters $[2^{2m-1}, 2m+1, 2^{2m-2}-2^{m-1}].$
Its weight enumerator is
$$1+2^{2m-2}z^{2^{2m-2}-2^{m-1}}+(3\cdot2^{2m-1}-2)z^{2^{2m-2}}+2^{2m-2}z^{2^{2m-2}+2^{m-1}}+z^{2^{2m-1}}.$$
The dual $\bar{{\cal C}}_{f_4,g_4}^{(2)\perp}$ is an optimal $[2^{2m-1}, 2^{2m-1}-2m-1, 4]$ binary code with respect to the sphere-packing bound.
\end{corollary}

\subsection{$f_5(x)=\Tr_{2^m/2}(x)$ and $g_5(y)=B(y)$ with $B(y)$ an Boolean bent function}\label{bent}
In this section, let $q=2$, $m$ be even and $f_5(x)=\Tr_{2^m/2}(x)$, $g_5(y)=B(y)$, where $B(y)$ is an Boolean bent functions from $\F_{2^m}$ to $\F_2$. For convenience, we abbreviate $\Tr_{2^m/2}(x)$ by $\Tr(x)$ for the rest of this subsection.

\begin{theorem}\label{sub_code5}
The code $\cC_{f_5,g_5}^{(2)}$ is a five-weight binary linear code with parameters $[2^{2m-1}+1, 2m+1, 2^{2m-2}- 2^{\frac{3m-4}{2}}]$ and
the weight distribution is given in Table \ref{tableC5q-1}.
The dual $\cC_{f_5,g_5}^{(2)\bot}$ is a $[2^{2m-1}+1,2^{2m-1}-2m, 3]$ binary code.
\end{theorem}

\begin{IEEEproof}
For $f_5(x)=\Tr(x)$, $g_5(y)=B(y)$, by the transitivity of trace functions, we have
$$\#D= 2^{2m-1} + \frac{1}{2} \sum\limits_{x \in \F_{2^m}}\chi(\Tr(x)) \sum\limits_{y \in \F_{2^m}}\chi(B(y))= 2^{2m-1}.$$
Then the code length of ${\cal C}_{f_5,g_5}^{(2)}$ is $n=2^{2m-1}+1.$
For any $a \in \F_2$, $(b,c) \in \F_{2^m}^2 \backslash (0,0)$,
  \[\begin{aligned}
\Upsilon_{a,b,c}&=(-1)^a \sum\limits_{x \in \F_{2^m}}\chi'\left((1+b)x\right) \sum\limits_{y \in \F_{2^m}}\chi\left(B(y)+ \Tr(cy)\right)\\
 &= \begin{cases}
       0  & \text { if } b\neq 1, \\
       (-1)^a2^m W_B(c)  & \text { if } b = 1,\\
       \end{cases}
   \end{aligned}\]
where $W_B(c)= \pm 2^{\frac{m}{2}}$ from the definition of Boolean Bent function.

Since $m$ is even, $\Tr(1)= m =0$. By Lemma \ref{weight}, for any codeword $\textbf{c}_{a,b,c}$ in $\cC_{f_5,g_5}^{(2)}$ we deduce that
$$
  wt(\textbf{c}_{a,b,c})=\begin{cases}
0 & \text { if } a=b=c=0,\\
2^{2m-1}  & \text { if } a\neq 0, b=c=0, \\
2^{2m-2}  & \text { if } b=0, c\neq0,\\
           & ~\text { or } b \notin \{0,1\}, \delta(b)=0, \\

2^{2m-2}+1  & \text { if } b\neq1,\delta(b)=1,\\

2^{2m-2}- 2^{\frac{3m-4}{2}} & \text { if }   b=1, a=0,  W_B(c)=2^{\frac{m}{2}},\\
                              & ~\text { or }   b=1, a=1,  W_B(c)=-2^{\frac{m}{2}},\\
2^{2m-2}+ 2^{\frac{3m-4}{2}} & \text { if }   b=1, a=0,  W_B(c)=-2^{\frac{m}{2}},\\
                              &~ \text { or }   b=1, a=1,  W_B(c)=2^{\frac{m}{2}},\\
    \end{cases}
$$

$$
=\begin{cases}
0 & \text { with 1 time,}\\
2^{2m-1}  & \text { with 1 time,}\\
2^{2m-2}  & \text { with $2^{2m}-2^{m+1}-2$ times,} \\
2^{2m-2}+1  & \text { with $2^{2m}$ times,}\\
2^{2m-2}- 2^{\frac{3m-4}{2}} & \text {  with $2^m$ times,}\\
2^{2m-2}+ 2^{\frac{3m-4}{2}} & \text {  with $2^m$ times,}\\
    \end{cases}
$$
where the last two numbers are determined because they have the same value. The dimension of $\cC_{f_5,g_5}^{(2)}$ is $2m+1$ as $A_0=1.$

Note that $\cC_{f_5,g_5}^{(2)\perp}$ is of length $2^{2m-1} + 1$ and dimension $2^{2m-1}-2m$. Similarly to the proof of Theorem \ref{sub_code3},
by Lemma \ref{weight} and the sphere-packing bound of codes, the minimal
distance of $\cC_{f_5,g_5}^{(2)\perp}$ satisfied $3 \leq  d_5^{(2)\perp} \leq 4$.
 By the first four Pless power moments, one can derive that $d_5^{(2)\perp} =3$.
\end{IEEEproof}

\begin{table}[h!]
  \begin{center}
    \caption{The weight distribution of $\cC_{f_5,g_5}^{(2)}$.}
    \begin{tabular}{c|c}
    \hline
      Weight & Multiplicity \\
      \hline
      0 & 1 \\
      $2^{2m-2}-2^{\frac{3m-4}{2}}$ & $2^{m}$ \\
      $2^{2m-2}$ & $2^{2m}-2^{m+1}-2$ \\
      $2^{2m-2}+1$ & $2^{2m}$ \\
      $2^{2m-2}+2^{\frac{3m-4}{2}}$ & $2^{m}$  \\
      $2^{2m-1}$ & 1 \\
       \hline
    \end{tabular}
    \label{tableC5q-1}
  \end{center}
\end{table}

We can directly get the following results from Lemma \ref{weight-2} and Theorem \ref{sub_code5}.

\begin{theorem}\label{punc_code5}
The linear code $\bar{{\cal C}}_{f_5,g_5}^{(2)}$ is a four-weight binary linear code with parameters $[2^{2m-1}, 2m+1, 2^{2m-2}-2^{\frac{3m-4}{2}}]$ and the weight distribution in Table \ref{tableC5q-2}.
The dual code $\bar{{\cal C}}_{f_5,g_5}^{(2)\perp}$ is an optimal $[2^{2m-1}, 2^{2m-1}-2m-1, 4]$ binary code with respect to the sphere-packing bound.
\end{theorem}

\begin{table}[h!]
  \begin{center}
    \caption{The weight distribution of $\bar{{\cal C}}_{f_5,g_5}^{(2)}$.}
    \begin{tabular}{c|c}
    \hline
      Weight & Multiplicity \\
      \hline
      0 & 1 \\
      $2^{2m-2}-2^{\frac{3m-4}{2}}$ & $2^{m}$ \\
      $2^{2m-2}$ & $2^{2m+1}-2^{m+1}-2$ \\
      $2^{2m-2}+2^{\frac{3m-4}{2}}$ & $2^{m}$  \\
      $2^{2m-1}$ & 1 \\
       \hline
    \end{tabular}
   \label{tableC5q-2}
  \end{center}
\end{table}

\subsection{$f_6(x)=B_1(x)$ and $g_6(y)=B_2(y)$ with $B_1$, $B_2$ are two different Boolean bent functions}\label{bent2}
In this section, let $q=2$, $m$ be even and $f_6(x)=B_1(x)$, $g_6(y)=B_2(y)$, where $B_1$, $B_2$ are two different Boolean bent functions from $\F_{2^m}$ to $\F_2$.

\begin{theorem}\label{sub_code6}
The code $\cC_{f_6,g_6}^{(2)}$ is a five-weight binary linear code with parameters $[2^{2m-1}+\frac{W}{2}+1, 2m+1, 2^{2m-2}+\frac{W}{4}-2^{m-2}]$, where $W=\pm 2^m$.
The weight distribution is given in Table \ref{tableC6q-1}.
The dual $\cC_{f_6,g_6}^{(2)\bot}$ is a $[2^{2m-1}+\frac{W}{2}+1, 2^{2m-1}+\frac{W}{2}2m, 3]$ binary code.
\end{theorem}

\begin{IEEEproof}
For $f_6(x)=B_1(x)$, $g_6(y)=B_2(y)$, by the transitivity of trace functions, we have
$$\#D= 2^{2m-1} + \frac{1}{2} \sum\limits_{x \in \F_{2^m}}\chi(B_1(x)) \sum\limits_{y \in \F_{2^m}}\chi(B_2(y))= 2^{2m-1}+\frac{W}{2},$$
where $W= W_{B_1}(0)W_{B_2}(0)=\pm 2^m$.
Then the code length of ${\cal C}_{f_6,g_6}^{(2)}$ is $n=2^{2m-1}+\frac{W}{2}+1.$
For any $a \in \F_2$, $(b,c) \in \F_{2^m}^2 \backslash (0,0)$,
  \[\begin{aligned}
\Upsilon_{a,b,c}&=(-1)^a \sum\limits_{x \in \F_{2^m}}(-1)^{B_1(x)+\Tr(bx)} \sum\limits_{y \in \F_{2^m}}(-1)^{B_2(y)+ \Tr(cy)}\\
 &= (-1)^a W',
   \end{aligned}\]
 where $W'=W_{B_1}(b)W_{B_2}(c)= \pm 2^{m}$.

 By Lemma \ref{weight}, for any codeword $\textbf{c}_{a,b,c}$ in $\cC_{f_6,g_6}^{(2)}$ we deduce that
  \[\begin{aligned}
  wt(\textbf{c}_{a,b,c})&=\begin{cases}
0 & \text { if } a=b=c=0,\\
2^{2m-1}+\frac{W}{2}  & \text { if } a\neq 0, b=c=0, \\
2^{2m-2}+\frac{W}{4}-2^{m-2}  & \text { if $b$ and $c$ are not all 0, } a=0, ~ W'=2^{m}, \delta(b)=0, \\
           & ~~~~~~~~~~~~~~~~~~~~~~~~~\text { or } a=1, ~ W'=-2^{m}, \delta(b)=0, \\

2^{2m-2}+\frac{W}{4}-2^{m-2}+1  & \text { if $b$ and $c$ are not all 0, } a=0, ~ W'=2^{m}, \delta(b)=1, \\
           & ~~~~~~~~~~~~~~~~~~~~~~~~~\text { or } a=1, ~ W'=-2^{m}, \delta(b)=1, \\

2^{2m-2}+\frac{W}{4}+2^{m-2}  & \text { if $b$ and $c$ are not all 0, } a=0, ~ W'=-2^{m}, \delta(b)=0, \\
           & ~~~~~~~~~~~~~~~~~~~~~~~~~\text { or } a=1, ~ W'=2^{m}, \delta(b)=0, \\

2^{2m-2}+\frac{W}{4}+2^{m-2}+1  & \text { if $b$ and $c$ are not all 0, } a=0, ~ W'=-2^{m}, \delta(b)=1, \\
           & ~~~~~~~~~~~~~~~~~~~~~~~~~\text { or } a=1, ~ W'=2^{m}, \delta(b)=1, \\
    \end{cases}\\
&=\begin{cases}
0 & \text { with 1 time,}\\
2^{2m-1}+\frac{W}{2}  & \text { with 1 time,}\\
2^{2m-2}+\frac{W}{4}-2^{m-2}  & \text { with $i_1$ times,} \\
2^{2m-2}+\frac{W}{4}-2^{m-2}+1  & \text { with $i_2$ times,}\\
2^{2m-2}+\frac{W}{4}+2^{m-2} & \text {  with $i_3$ times,}\\
2^{2m-2}+\frac{W}{4}+2^{m-2}+1 & \text {  with $i_4$ times.}\\
    \end{cases}
 \end{aligned}\]
It's obvious that $i_1=i_3$, $i_2=i_4$ and the dimension of $\cC_{f_6,g_6}^{(q)}$ is $2m+1$ as $A_0=1.$
Since any two columns of matrix $G_{f_6,g_6}$ are linearly independent. It follows from Lemma \ref{distance} that the minimal distance of $\cC_{f_6,g_6}^{(2)\perp}$ satisfies $d_6^{(2)\perp} \geq 3$. 
Then by the first three Pless power moments, one can derive that $i_1=i_3=2^{2m-1}-1$ and $i_2=i_4= 2^{2m-1}$.

Note that $\cC_{f_6,g_6}^{(2)\perp}$ is of length $2^{2m-1} + 1$ and dimension $2^{2m-1}-2m$. Similar to the proof of Theorem \ref{sub_code3},
by Lemma \ref{weight} and the sphere-packing bound of codes, we have the minimal
distance of $\cC_{f_6,g_6}^{(2)\perp}$ satisfied $3 \leq  d_6^{(2)\perp} \leq 4$.
 By the first four Pless power moments, one can derive that $d_6^{(2)\perp} =3$.
\end{IEEEproof}

\begin{table}[h!]
  \begin{center}
    \caption{The weight distribution of $\cC_{f_6,g_6}^{(2)}$.}
    \begin{tabular}{c|c}
    \hline
      Weight & Multiplicity \\
      \hline
      0 & 1 \\
      $2^{2m-2}+\frac{W}{4}-2^{m-2}$ & $2^{2m-1}-1$ \\
      $2^{2m-2}+\frac{W}{4}-2^{m-2}+1$ & $2^{2m-1}$\\
      $2^{2m-2}+\frac{W}{4}+2^{m-2}$ & $2^{2m-1}-1$ \\
      $2^{2m-2}+\frac{W}{4}+2^{m-2}+1$ & $2^{2m-1}$  \\
      $2^{2m-1}+\frac{W}{2}$ & 1 \\
       \hline
    \end{tabular}
    \label{tableC6q-1}
  \end{center}
\end{table}

We can directly get the following results from Lemma \ref{weight-2} and Theorem \ref{sub_code6}.

\begin{theorem}\label{punc_code6}
The code $\bar{{\cal C}}_{f_6,g_6}^{(2)}$ is a three-weight binary linear code with parameters $[2^{2m-1}+\frac{W}{2}, 2m+1, 2^{2m-2}+\frac{W}{4}-2^{m-2}]$ and the weight distribution in Table \ref{tableC6q-2}.
The dual code $\bar{{\cal C}}_{f_6,g_6}^{(2)\perp}$ is an optimal $[2^{2m-1}+\frac{W}{2}, 2^{2m-1}+\frac{W}{2}-2m-1, 4]$ binary code with respect to the sphere-packing bound.
\end{theorem}

\begin{table}[h!]
  \begin{center}
    \caption{The weight distribution of $\bar{{\cal C}}_{f_6,g_6}^{(2)}$.}
    \begin{tabular}{c|c}
    \hline
      Weight & Multiplicity \\
      \hline
      0 & 1 \\
      $2^{2m-2}+\frac{W}{4}-2^{m-2}$ & $2^{2m}-1$ \\
      $2^{2m-2}+\frac{W}{4}+2^{m-2}$ & $2^{2m}-1$  \\
      $2^{2m-1}+\frac{W}{2}$ & 1 \\
       \hline
    \end{tabular}
   \label{tableC6q-2}
  \end{center}
\end{table}

\section{Application in $t$-Designs}\label{design}
Let $\kappa$ and $n$ be positive integers such that $1 \leq \kappa \leq n$.  Let $\cP$ be a set of $n$ elements and let $\cB$ be a set of $\kappa$-subsets of $\cP$.
Let $t$ be a positive integer with $t \leq \kappa$. The pair $\mathbb{D} = (\cP,\cB)$ is an \emph{incidence structure}, where the incidence relation is the set membership.
The incidence structure $\mathbb{D} = (\cP,\cB)$ is called a $t-(n, \kappa, \lambda)$ design, or simply $t$-design, if every $t$-subset of $\cP$ is contained in exactly $\lambda$ elements of $\cB$. The elements of $\cP$ are called points, and those of $\cB$ are referred to as blocks.
A $ t-$design is said to be \emph{simple} if $\cB$ does not contain any repeated blocks.
A $t-(n,\kappa,\lambda)$ design is called a \emph{Steiner system} if $t \geq 2$ and $\lambda = 1$, and is denoted by $S(t,\kappa, n)$.
Let $\mathfrak{b}$ denote the number of blocks in $\cB$. The parameters of a $t-(n,\kappa,\lambda)$ design satisfy the following equation:
$$\binom{n}{t}\lambda = \binom{\kappa}{t}\mathfrak{b}.$$

A construction of $t$-designs with linear codes can be obtained as follows. Let $\cC$ be an $[n, k, d]$ linear code over $\F_q$. 
Let the coordinates of a codeword of $\cC$ be indexed by $(0,1, \cdots, n-1)$ and define $\cP(\cC)$ = $\{0,1, \cdots ,n-1\}$. For a codeword $\textbf{c} = (c_0, c_1, \cdots, c_{n-1}) \in \cC$, the support of $\textbf{c}$ is defined by $\mathrm{suppt}(\textbf{c})=\{i : c_i \neq 0, i \in \cP(\cC)\}.$
Let $\cB_{\kappa}(\cC)$ denote the set of the supports of all codewords with Hamming weight $\kappa$ in $\cC$ without repeated blocks.
If the incidence structure $\mathbb{D}_{\kappa} = (\cP(\cC),\cB_{\kappa}(\cC))$ is a $t-(n,\kappa,\lambda)$ design for some positive integers $t$ and $\lambda$, where $1 \leq \kappa \leq n$ and $A_{\kappa} \neq 0$, we say that the code $\cC$ holds a $t$-design or the supports of the codewords of weight $\kappa$ in $\cC$ hold a $t$-design.

The following theorem, which Assumus and Mattson developed in \cite{Assmus1969}, provides a necessary condition for a linear code and its dual to hold simple $ t-$designs.

\begin{theorem}[Assumus-Mattson Theorem]\label{AM} Let $\cC$ be an $[n, k, d]$ code over $\F_q$. Let $d^{\perp}$ denote the minimum distance of $\cC^{\perp}$. Let $w$ be the largest integer satisfying $w \leq n$ and
$$
w-\left\lfloor\frac{w+q-2}{q-1}\right\rfloor<d .
$$
Define $w^{\perp}$ analogously using $d^{\perp}$. Let $\left(A_i\right)_{i=0}^n$ and $\left(A_i^{\perp}\right)_{i=0}^n$ denote the weight distribution of $\cC$ and $\cC^{\perp}$, respectively. Fix a positive integer $t$ with $t<d$, and let $s$ be the number of $i$ with $A_i^{\perp} \neq 0$ for $1 \leq i \leq n-t$. Suppose $s \leq d-t$. Then
\begin{itemize}
  \item the codewords of weight $i$ in $\cC$ hold a simple $t$-design provided $A_i \neq 0$ and $d \leq i \leq w$, and
  \item the codewords of weight $i$ in $\cC^{\perp}$ hold a simple $t$-design provided $A_i^{\perp} \neq 0$ and $d^{\perp} \leq i \leq \min \left\{n-t, w^{\perp}\right\}.$
\end{itemize}
\end{theorem}

\begin{theorem}\label{design1}
The codewords of Hamming weight $q^3-q$ in the code $\cC_{f_1,g_1}$ hold a $2-(q^3+1, q^3-q, (q -1)(q^3 -
q - 1))$ design. The complementary design of this design is a $2-(q^3+1, q+1, 1)$ design, i.e., a Steiner system $S(2, q+1, q^3+1)$.
The minimum weight codewords in $\cC_{f_1,g_1}^{\bot}$ hold a $2-(q^3+1, 3, \lambda)$ design, where
$$\lambda = \frac{6 A_3^{\bot}}{q^3(q^2-1)(q^3+1)}$$
and $A_3^{\bot}$ denotes the number of codewords of weight 3 in $\cC_{f_1,g_1}^{\bot}$.

\end{theorem}
\begin{IEEEproof}
The desired conclusions follow from Theorem \ref{C_1} and the Assmus-Mattson Theorem. In addition, the quantity
$A_3^{\bot}$ can be computed with the MacWilliams identity and the weight enumerator of $\cC_{f_1,g_1}$.
\end{IEEEproof}

\begin{theorem}\label{design2}
Let $m \geq 2$. The codewords of Hamming weight $2^{2m-2}$ in the code $\bar{\cC}_{f_2,g_2}^{(2)}$ hold a $3-(2^{2m-1}, 2^{2m-2},$ $ 2^{2m-3}-1)$ design. For even positive integer $\kappa$ with $4 \leq \kappa \leq 2^{2m-1}-4$, the codewords of Hamming weight $\kappa$ in the code $\bar{\cC}_{f_2,g_2}^{(2)\bot}$ hold a $3$-design.

\end{theorem}
\begin{IEEEproof}
By Remark \ref{re4}, $\bar{\cC}_{f_2,g_2}^{(2)}$ is binary Reed-Muller code $\mathcal{RM}(1,2m-1)$. It is
known the binary Reed-Muller code $\mathcal{RM}(1,2m-1)$ hold 3-designs \cite{Ding2018}. The desired conclusions then follow.
\end{IEEEproof}

\begin{theorem}\label{design3}
Let $q=2$. Then the codewords of Hamming weight $2^{2m-2}+\frac{W}{4}-2^{m-2}$ or $2^{2m-2}+\frac{W}{4}+2^{m-2}$, where $W=\pm 2^m$, in the code $\bar{\cC}_{f_6,g_6}^{(2)}$ hold a $2$-design. 
For any $4 \leq \kappa \leq 2^{2m-1}+\frac{W}{2}$, the codewords of Hamming weight $\kappa$ in the code $\bar{\cC}_{f_6,g_6}^{(2)\bot}$ hold a $2$-design.

\end{theorem}
\begin{IEEEproof}
The desired conclusions follow from Theorem \ref{punc_code6} and the Assmus-Mattson Theorem.
\end{IEEEproof}

\section{Concluding Remarks}\label{conclusion}


The main contributions of this paper are the following:
\begin{itemize}
 \item A family of $[q^3+1, 3, q^3-q]$ two-weight linear code $\cC_{f_1,g_1}$ over $\F_{q^2}$ meeting the Griesmer bound was derived.
     We determined its weight enumerator and the parameters of its dual.
     It was shown that the dual code is an almost MDS code, and when $q=2$, $\cC_{f_1,g_1}$ is a quaternary Hermitian self-orthogonal code (see Theorem \ref{C_1}, Remark \ref{remark2}).


 \item A family $[q^{2m-1}+1, 3]$ linear codes with four or five weights over $\F_{q^m}$ for any positive integer $m$ was presented.
     We derived its weight enumerator and the parameters of its dual.
     It was shown that its minimal distance depends on the parities of $q$ and $m$, and its dual code is an almost MDS codes (see Theorem \ref{C_2}).

 \item The $q$-ary subfield codes and their punctured code for the above two families of few-weight linear codes were investigated.
     We determined their weight distribution and the parameters of their dual.
     It was shown that some of the resultant codes are optimal, and some have the best-known parameters (see Theorems \ref{sub_code1}, \ref{punc_code1}, \ref{sub_code2}, \ref{punc_code2} and Examples \ref{ex_sub_C1}, \ref{ex_punc_C1}, \ref{ex_sub_C2}, \ref{ex_punc_C2}).

 \item A family of $[2^{4m-2}, 2m+1, 2^{4m-3}]$ quaternary Hermitian self-dual code was obtained, where $m\geq2$ (see Remark \ref{re4}).

 \item Seven families of few-weight binary linear codes were presented by employing almost bent and Boolean bent functions. Most of their dual codes are optimal with respect to the sphere packing bound (see theorems \ref{sub_code3}-\ref{punc_code6}).

 \item Several infinite families of $2$-designs or $3$-designs were constructed from three families of linear codes given in this paper (see Theorems \ref{design1}, \ref{design2} and \ref{design3}).
\end{itemize}

The trace, norm, almost bent, and Boolean bent functions were used in this paper. By selecting, additionally, suitable functions $f(x)$ and $g(x)$ to construct the generator matrix $G_{f,g}$, it would be possible to get more few-weight codes with optimal parameters.

\section*{Acknowledgments}

The authors warmly thank Prof. Zhengchun Zhou for his interesting discussion and several exciting suggestions during this work.

\end{document}